\documentclass[prb,aps,amssymb,twocolumn,superscriptaddress,notitlepage]{revtex4-1}
\usepackage{amsmath}
\usepackage{amssymb}
\usepackage{amsthm}
\usepackage{amsfonts}
\usepackage{listings}
\usepackage{enumerate}
\usepackage{latexsym}
\usepackage{psfrag}
\usepackage{bm}
\usepackage{graphicx}
\usepackage[caption=false]{subfig}
\usepackage{blkarray}
\usepackage{array}
\usepackage{color}
\usepackage[normalem]{ulem}
\usepackage{hyperref}

\makeatletter
\DeclareRobustCommand{\element}[1]{\@element#1\@nil}
\def\@element#1#2\@nil{%
  #1%
  \if\relax#2\relax\else\MakeLowercase{#2}\fi}
\makeatother

\newcolumntype{L}{>{$}l<{$}}

\newtheorem*{defn*}{Definition}

\newcommand{\half}{\frac{1}{2}}

\begin{document}
\title{Disconnected Elementary Band Representations, Fragile Topology, and Wilson Loops as Topological Indices: An Example on the Triangular Lattice}
\author{Barry Bradlyn}
\affiliation{Department of Physics and Institute for Condensed Matter Theory, University of Illinois at Urbana-Champaign, Urbana, IL, 61801-3080, USA}
\affiliation{Donostia International Physics Center, P. Manuel de Lardizabal 4, 20018 Donostia-San Sebasti\'{a}n, Spain}
\author{Zhijun Wang}
\affiliation{Beijing National Laboratory for Condensed Matter Physics and Institute of Physics, Chinese Academy of Sciences, Beijing, China}
\affiliation{Department of Physics, Princeton University, Princeton, New Jersey 08544, USA}
\author{Jennifer Cano}
\affiliation{Department of Physics and Astronomy, Stony Brook University, Stony Brook, New York 11974, USA}
\affiliation{Center for Computational Quantum Physics, The Flatiron Institute, New York, New York 10010, USA}
\author{B. Andrei Bernevig}
\affiliation{Department of Physics, Princeton University, Princeton, New Jersey 08544, USA}
\affiliation{
Dahlem Center for Complex Quantum Systems and Fachbereich Physik,
Freie Universit{\"a}t Berlin, Arnimallee 14, 14195 Berlin, Germany
	}
\affiliation{
Max Planck Institute of Microstructure Physics, 
06120 Halle, Germany
}
\date{\today}
\begin{abstract}
In this work, we examine the topological phases that can arise in triangular lattices with disconnected elementary band representations. We show that, although these phases may be ``fragile'' with respect to the addition of extra bands, their topological properties are manifest in certain nontrivial holonomies (Wilson loops) in the space of nontrivial bands. {We introduce an eigenvalue index for fragile topology, and we show how a nontrivial value of this index manifests as the winding of a hexagonal Wilson loop; this remains true} even in the absence of time-reversal or sixfold rotational symmetry. Additionally, when time-reversal and twofold rotational symmetry are present, we show directly that there is a protected nontrivial winding in more conventional Wilson loops. Crucially, we emphasize that these Wilson loops cannot change without closing a gap \emph{to the nontrivial bands}. By studying the entanglement spectrum for the fragile bands, we comment on the relationship between fragile topology and the ``obstructed atomic limit'' of Ref.~\onlinecite{NaturePaper}. We conclude with some perspectives on topological matter beyond the K-theory classification.
\end{abstract}
\maketitle

\section{introduction}

A recent work (Ref.~\onlinecite{NaturePaper}) developed a topological band theory predicated on the notion of elementary band representations (EBRs), which are the fundamental building blocks of atomic-limit band structures.\cite{Zak1980,Zak1981,aris18-1} Complementing and including other theories classifying topological phases \cite{Po2017,Shiozaki2017,Freed2013,Kruthoff2016}, this theory provides a clear map from orbitals in real space to topology in momentum space; as such, using this theory we were able (uniquely amongst classifying theories) to predictively construct and tabulate all the types of bands appearing in the 230 space groups,\cite{GroupTheoryPaper,GraphTheoryPaper} with all possible orbitals at all Wyckoff positions, and to predict and identify large classes of new topological materials. {Additionally, these EBR tables\cite{GroupTheoryPaper,GraphTheoryPaper} were used by various authors\cite{song2017,bigmaterials-china} to deduce explicit forms for all symmetry-indicated strong topological indices}. Using this theory, it was proved that if a system consisted of disconnected bands separated by a gap in momentum space, and if all eigenfunctions taken together transform as a single EBR, then these bands cannot be topologically trivial\cite{NaturePaper,EBRTheoryPaper,aris18-2}.

Building on this notion, Po et al. have constructed\cite{comment} a model in wallpaper group $p6mm1'$ (space group $183$ with time-reversal ($T$) symmetry)\footnote{Because we will be considering in this work models both with and without time-reversal symmetry, we will follow Ref.~\onlinecite{Cracknell} and place a ``$1'$'' after a space group symbol when we consider time-reversal to be a symmetry of the system.} which realizes one of the disconnected elementary band representation first explicitly presented in Ref.~\onlinecite{NaturePaper}. Their model consists of spinful $p_z$ orbitals on the $2b$ Wyckoff position (honeycomb lattice site, c.f.~Fig.~\ref{fig:realspacemodel}), similar to graphene. The four bands taken together transform as a (physically, since it is time-reversal symmetric) elementary band representation (PEBR). In Ref.~\onlinecite{NaturePaper}, it was shown that the disconnected realizations of {this} elementary band representation heralds the possibility of the existence of the $\mathbb{Z}_2$ nontrivial phase of the Kane-Mele model with Rashba spin-orbit coupling.  However, by including sufficiently long range hoppings, the model of Po et al.~realizes a different phase, where, although one band -- which, in their model, has the highest energy --  has no symmetric, localized Wannier description, the authors are able to construct symmetric, localized Wannier functions for the other, lower energy, band, centered at the origin of the unit cell, the $1a$ position (although they have no weight at the $1a$ position).  Furthermore, \emph{by adding} two ancillary orbitals on the $1a$ position, the authors are able to deform the lower energy band of their model to an atomic limit, where the Wannier functions for the lowest two bands {are not only centered on the $1a$ Wyckoff position, but also vanish on all other sites (i.e., they are atomic-limit $s$ orbitals at the $1a$ position)}. This and related models have been recently explored further in Ref.~\onlinecite{Slager2018} {(which appeared while this work was in preparation)}, where the authors showed that, with twofold rotational symmetry and time-reversal symmetry, there existed a Wilson loop invariant which can be used to detect that an isolated set of two bands does not admit localized, symmmetric Wannier functions.

In Ref.~\onlinecite{Fragile2017}, we addressed many of these issues through an examination of several disconnected elementary band representations without spin-orbit coupling. We showed that in all cases considered there existed a topological group of bands, i.e. a set of bands which do not admit a description in terms of localized, symmetric Wannier functions. Furthermore, we showed that this Wannier obstruction could be detected by the nontrivial winding of an appropriately defined Wilson loop within the space of topological bands. Because the winding of these Wilson loops cannot change without closing a gap to the topological bands, the winding number serves as an index of fragile topology (c.f.~Ref.~\onlinecite{Slager2018}). We emphasized that even these ``fragile'' invariants correspond to in-principle measurable phenomena, as they are properties of an isolated set of bands; observable consequences of fragile topology have recently been studied in Refs.~\onlinecite{wieder2018axion,ahn2018higher,benalcazar2018quantization}

{The results of Refs.~\onlinecite{NaturePaper}, \onlinecite{EBRTheoryPaper}, \onlinecite{Fragile2017}, and \onlinecite{Slager2018} are all consistent: \emph{a disconnected elementary band representation cannot be trivial}. 
In this paper, we further explore the models and theory of fragile topology, and show the stability of fragile phases through Wilson loop calculations.}  
We first focus on the specific case of the model of Po et al.~in Sec.~\ref{sec:model}, and
show that this model in fact is an explicit demonstration how the theory of band representations\cite{NaturePaper,EBRTheoryPaper} is able to capture topological behavior beyond the usual (stable) K-theoretic classification. Next, we show that the topologically nontrivial bands in the model remain nontrivial upon relaxing first time reversal and then twofold rotational symmetry. We highlight that with these relaxed symmetries the conventional Wilson loop of Ref.~\onlinecite{Slager2018} and Section~\ref{sec:model} is no longer a good indicator of nontrivial topology. {It would then seem as if we had constructed a disconnected EBR with no nontrivial Wilson loops; however, we show that this is not the case.}  We introduce a new, hexagonal Wilson loop which can detect the presence of a split elementary band representation using only threefold rotation and mirror symmetry, thus highlighting the point that one must consider \emph{all} possible Wilson loops (and, more generally, nested Wilson loops and the analytic structure of symmetry matrices\cite{wieder2018axion,benalcazar2018quantization,ahn2018higher}) in order to deduce that a group of bands is topologically trivial. The winding number of this hexagonal Wilson loop serves as an index of fragile topology even in the absence of time-reversal symmetry. Furthermore, we show how the presence of this non-trivial winding can be determined from the symmetry of the occupied bands. We then conclude in Sec.~\ref{sec:conclusion} by presenting how these observations fit into a broader view on topological band theory where, by definition, a set of separate bands is topological if at least one of the separate bands in the system cannot be described by a band representation as presented in  [\onlinecite{NaturePaper}].  We advocate for an expansion of the notion of ``topological phase'' beyond the conventional definition in terms of a topologically nontrivial projector onto the space of \emph{all occupied bands}; rather, we propose that it is relevant to examine the topology of \emph{isolated groups of bands} (i.e. projectors). The theory of topological quantum chemistry\cite{NaturePaper} (TQC) is uniquely suited to all these tasks.

\section{$C_2T$-symmetric model in the framework of TQC}\label{sec:model}

\subsection{Four Band Model}
In Ref.~\onlinecite{comment}, the authors give the full details of their model, which we summarize in Appendix~\ref{app:models}. In its simplest form, the model consists of spinful $p_z$ orbitals centered on both sites of a honeycomb lattice, the $2b$ position of space group $p6mm1'$. By including long-distance hoppings and exotic spin-orbit interactions, this model realizes a disconnected EBR (first discovered in Ref.~\onlinecite{NaturePaper}) in a topological phase distinct from the usual Kane-Mele model (it is $\mathbb{Z}_2$ trivial.) The lowest set of bands in this model span the little group representations $\bar{\Gamma}_9$ at the $\Gamma$ point, and $\bar{K}_6$ at the $K$ point, which are the same irreps as the band representation $(\bar{E}_1\uparrow G)_{1a}$ induced from $s$ orbitals at the center of the unit cell. The authors find symmetric, localized Wannier functions for this set of bands (which can be made to have lowest energy - valence). The other set of bands (which can be chosen to be the conduction bands) admits no description in terms of localized orbitals -- the projection operator onto these eigenstates is topologically nontrivial -- as pointed out by us in Ref.~\onlinecite{EBRTheoryPaper}. This can be seen from the little group representations for this group of bands, which are $\bar{\Gamma}_8$ at the $\Gamma$ point, and $\bar{K}_4\oplus\bar{K}_5$ at the $K$ point. Consulting the Bilbao Crystallographic Server\cite{GroupTheoryPaper}, we note that these little group representations cannot be obtained from any two-band EBR, but rather arise only as the formal difference of band representations\cite{Po2017,Fragile2017}. Thus, we confirm that these bands, taken in isolation, admit no localized, symmetric Wannier description. We emphasize here that while an examination of the little group irreps is a sufficient condition to deduce that a group of bands is topologically nontrivial, it is not a necessary condition; the triviality of the ``valence'' bands in this model must be deduced by other means, as was done in Ref.~\onlinecite{comment} {through an explicit construction of localized symmetric Wannier functions}.

This is consistent with the general proof given in Ref.~\onlinecite{EBRTheoryPaper}, where we showed that if an elementary band representation is disconnected into two (or more) groups of bands, \emph{at least one} of these groups must be topologically nontrivial (the situation when only one, rather than two bands are nontrivial, was not explicitly acknowledged in Ref.~\onlinecite{NaturePaper}). 
{The theory of EBRs thus gives a unified description of both strong topology\cite{Po2017,song2017} and fragile topology\cite{comment,Fragile2017,Slager2018}.}
The decomposition of an EBR into a topological band along with a band admitting a localized Wannier description demonstrates that, unlike previous methods, we are able to capture nontrivial topology beyond the K-theoretic classification. We point out that our EBR classification includes, as a subset, the symmetry index classification of Ref.~\onlinecite{Po2017}, obtained by modding out by the addition (and appropriately defined\cite{Freed2013} subtraction) of EBRs.  However, it also includes the cases that cannot be obtained by the definition of Ref.~\onlinecite{Po2017} and by the K-theory classification.
\begin{figure}[t]
\subfloat[]{
	\includegraphics[width=0.2\textwidth]{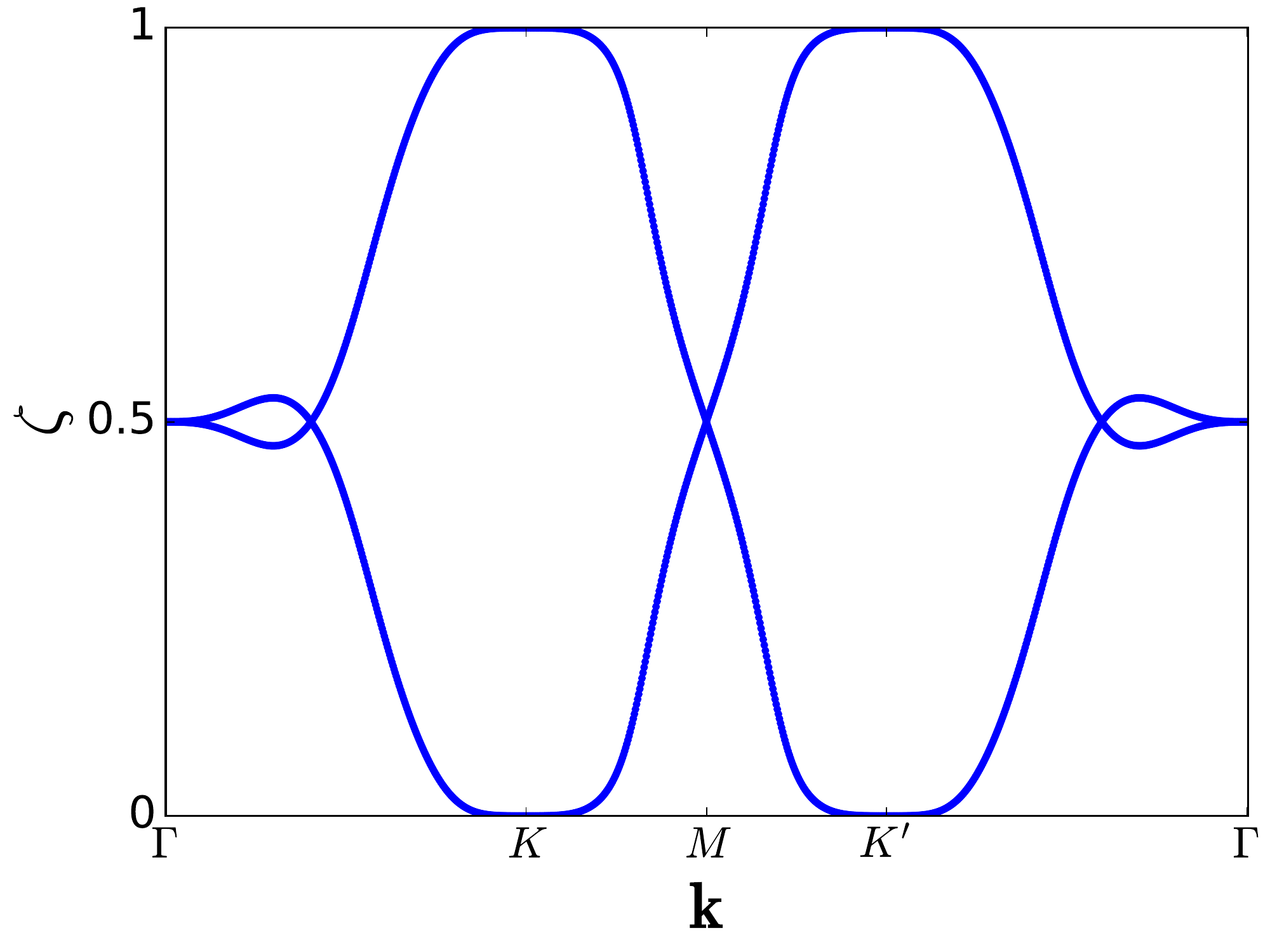}\label{fig:0muentspec}
}
\subfloat[]{
	\includegraphics[width=0.2\textwidth]{mu0ent.pdf}\label{fig:0muentspec2}
}\qquad
\subfloat[]{
	\includegraphics[width=0.2\textwidth]{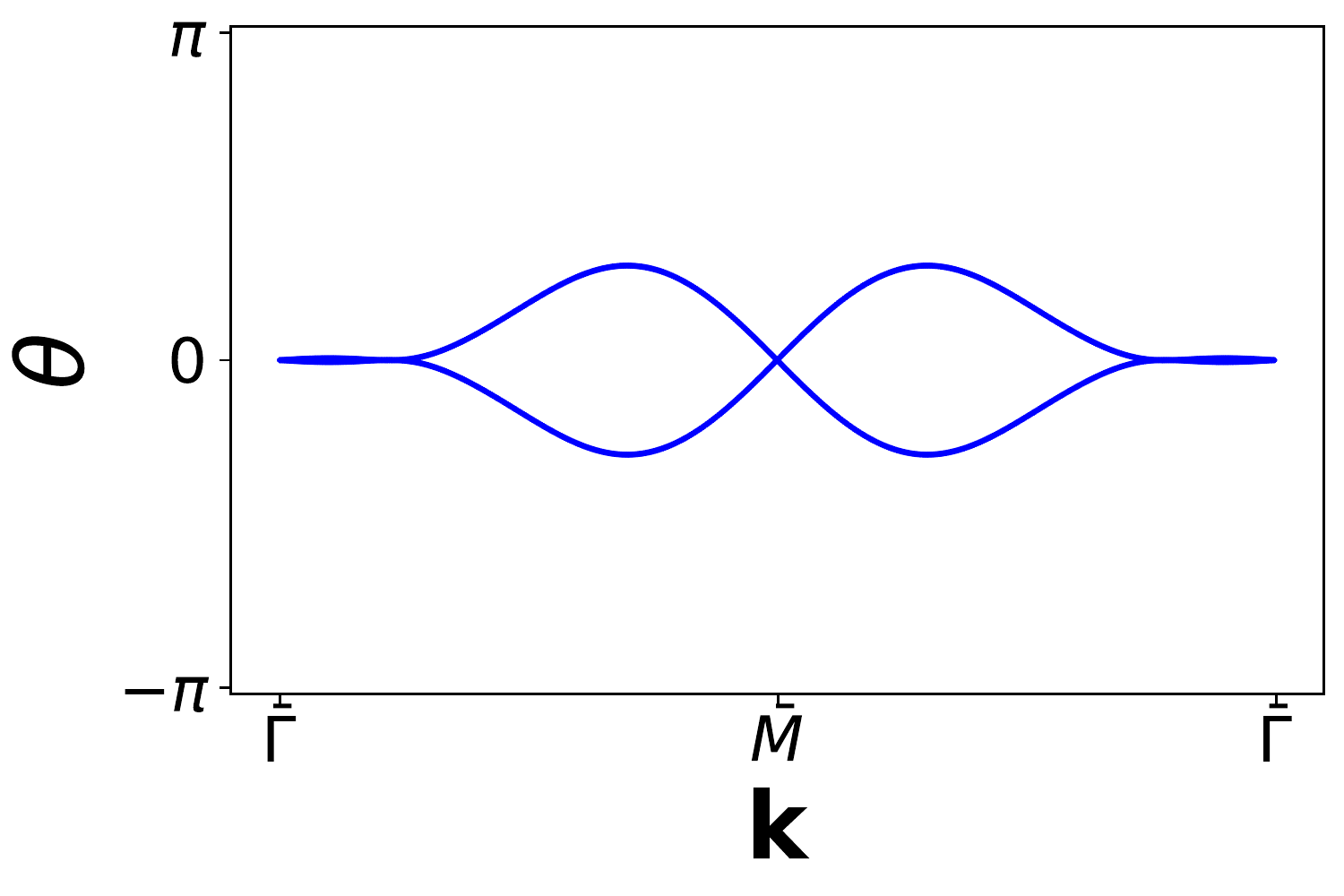}\label{fig:0muwilsonvalence}
}
\subfloat[]{
	\includegraphics[width=0.2\textwidth]{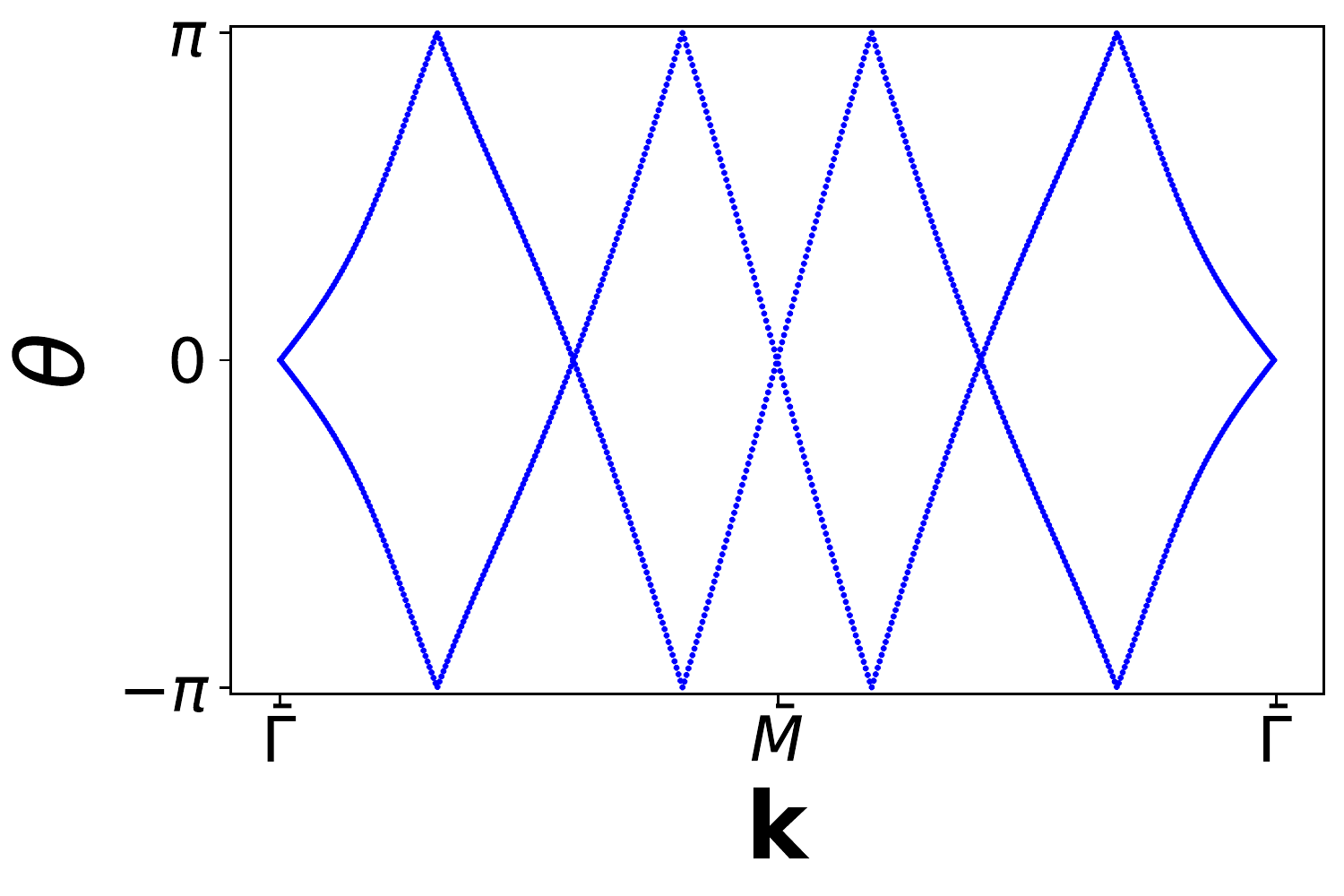}\label{fig:0muwilsoncond}
}
\caption{(a,b) Sublattice-resolved entanglement spectrum {(as defined in Eq.~\ref{eq:entregiondef}) for the higher (a) and lower (b) energy bands that comprise the disconnected EBR in the model Eq.~(\ref{eq:hamgen1}) when $\mu=0$.} $\zeta$ are the eigenvalues of the correlation matrix in the lowest two bands, restricted to the $A$ sublattice. The fact that $\zeta$ is gapless and extends from $0$ (Bloch functions have zero weight on the A sublattice) to $1$ (wavefunctions are entirely contained on the A sublattice) indicates that these bands cannot be continued to an unobstructed atomic limit. 
(c) Wilson loop of the lowest energy bands at $\mu=0$. The Wilson loop was evaluated in the $\mathbf{g}_2$ reciprocal lattice direction, and plotted as a function of the basepoint in the $\mathbf{g}_1$ direction. 
(d) Wilson loop of the higher energy bands in the disconnected EBR at $\mu=0$. The Wilson loop was evaluated in the $\mathbf{g}_2$ reciprocal lattice direction, and plotted as a function of the basepoint in the $\mathbf{g}_1$ direction. The winding of the Wilson loops shows that these bands do not have a localized Wannier description.
}
\end{figure}

As pointed out in Ref.~\onlinecite{Slager2018}, the topology of bands in this model can be diagnosed by considering Wilson loops
\begin{equation}
W^{mn}_{\mathbf{g}_1}(k_2)=\langle u^m_{\mathbf{g}_1+k_2\mathbf{g}_2}|\prod_{\mathbf{k}}^{k_2\mathbf{g}_2+\mathbf{g}_1\leftarrow k_2\mathbf{g}_2 }P(\mathbf{k})|u^n_{k_2\mathbf{g}_2}\rangle\label{eq:wilsondef}
\end{equation}
 oriented along the reciprocal lattice directions. The path of this loop is shown in Fig.~\ref{figstrtloop}. Here $P(\mathbf{k})$ is a projector onto a set of two bands. We first consider the case with twofold rotation ($C_2$) and time-reversal ($T$) symmetry. The Hamiltonian is given explicitly in Appendix~\ref{app:models}.  In Fig.~\ref{fig:0muwilsoncond} we plot the Wilson loop in the direction of the reciprocal lattice vector $\mathbf{g}_1$, plotted as a function of momentum in the $\mathbf{g}_2$ direction for the topological bands, which exhibit nontrivial winding. In Appendix~\ref{app:c2tproof}, we present a simple, direct proof that the crossings in these Wilson loops are protected by $C_2T$ symmetry, complementing the higher-level arguments of Ref.~\onlinecite{Slager2018}. These crossings are protected for \emph{any} two-band Wilson loop, irrespective of the total number of bands in the system; we will revisit this in Sec.\ref{subsec:6bands}. Because the winding of this Wilson loop is nontrivial, we deduce that these bands do not admit a description in terms of localized, symmetric Wannier functions, i.e. the bands are topological. We also show the same Wilson loop for the lower energy bands, which is trivial.

Concomitantly with the fact that the highest-energy band cannot be written in terms of exponentially localized symmetric Wannier states, the entanglement spectrum of the gapped valence bands is nontrivial. To exemplify this, we compute the entanglement spectrum using an entanglement region that includes all $A$ sublattice sites, {and excludes all $B$ sublattice sites.\cite{Legner2013} This entanglement region can be defined as the image of the projector
\begin{equation}
P_A=\sum_{\mathbf{R}\sigma}|\phi_{A\mathbf{R}\sigma}\rangle\langle\phi_{A\mathbf{R}\sigma}|,\label{eq:entregiondef}
\end{equation}
where $\phi_{A\mathbf{R}\sigma}$ is the tight binding basis function for the states on sublattice $A$ in unit cell $\mathbf{R}$ with spin $\sigma$.} (beyond the tight-binding limit, we can use instead the orthogonalized L{\"o}wdin orbitals\cite{Lowdin1950} centered on sublattice $A$, or even all functions supported on the $C_3$ symmetric half of the unit cell containing the $A$ sublattice) As shown in Fig.~\ref{fig:0muentspec}, the entanglement spectrum for this cut is gapless as a function of momentum, indicating that the two bands are inextricably linked. In the four-band model with orbitals only at the honeycomb lattice sites, the nontriviality of the entanglement spectrum follows immediately from a consideration of the $C_3$ eigenfunctions at high-symmetry points. In particular, it can be shown\footnote{See for instance Eq.~(S43) of Ref.~\onlinecite{NaturePaper}, and also Footnote $34$ of Ref.~\onlinecite{Soluyanov2011}} by directly examining the basis functions for the little group representations that for a four band model with $s$ or $p_z$ orbitals at the $2b$ position of wallpaper group $p6mm1'$, the $A$-sublattice entanglement spectrum for two bands with little group representions $(\bar{\Gamma_9}, \bar{K}_6)$ is pinned to $\zeta(\Gamma)=(1/2,1/2)$ and $\zeta(K)=(0,1)$. This shows that in the four-band model, {the valence band states cannot be localized entirely on one sublattice. Thus, we deduce that the Wannierizable (i.e.~describable in terms of a set of exponentially localized, symmetric Wannier functions)} valence bands form an obstructed atomic limit (i.e.~the Wannier functions are centered on a different Wyckoff position than the basis orbitals). It is precisely because the Wannier functions are not centered on the honeycomb lattice sites that the entanglement spectrun is nontrivial.

Taken together, these observations imply that the only way to deform the model to a limit where the valence band Wannier functions are indistinguishable from atomic orbitals without closing the gap is to extend the Hilbert space to include more than a single EBR. However, this process of adding states is not reversible: once we add extra states to allow us to disentangle the two bands in this disconnected EBR, there is no way to then remove the extra states without changing the electron filling, or the topology. Additionally, and as we will show in the next section, the model will always have at least one band not obtainable from exponentially localized Wannier orbitals.

\subsection{Six Band Model in the framework of TQC}\label{subsec:6bands}

We now extend our model by coupling to two ancillary states {(with Hilbert space on the $1a$ Wyckoff position at the center of the unit cell)}, giving six bands total. It was shown in Ref.~\onlinecite{comment} that by adding two additional states (one per spin) per unit cell centered at the $1a$ Wyckoff position, it is possible to construct a homotopy from the new,  Hilbert space enhanced, model to one in which the lowest two bands have atomic-limit Wannier functions supported entirely at the $1a$ position. Letting $H_0$ be the Hamiltonian of the original four-orbital model, Ref.~\onlinecite{comment} constructed (up to an overall scale factor)
\begin{equation}
H(\mu)=H_0\oplus\mathbf{0}_{2\times2}+\sin(\pi \mu)
H_c+\mathbf{0}_{4\times4} \oplus 36(1-2\mu)\mathbb{I}_{2\times 2},
\label{eq:model}
\end{equation}
where $H_c$ couples the original orbitals to the ancillary orbitals at the $1a$ position and the last term is a chemical potential for the ancillary orbitals.

\begin{figure}[h]
\subfloat[]{
	\includegraphics[width=0.2\textwidth]{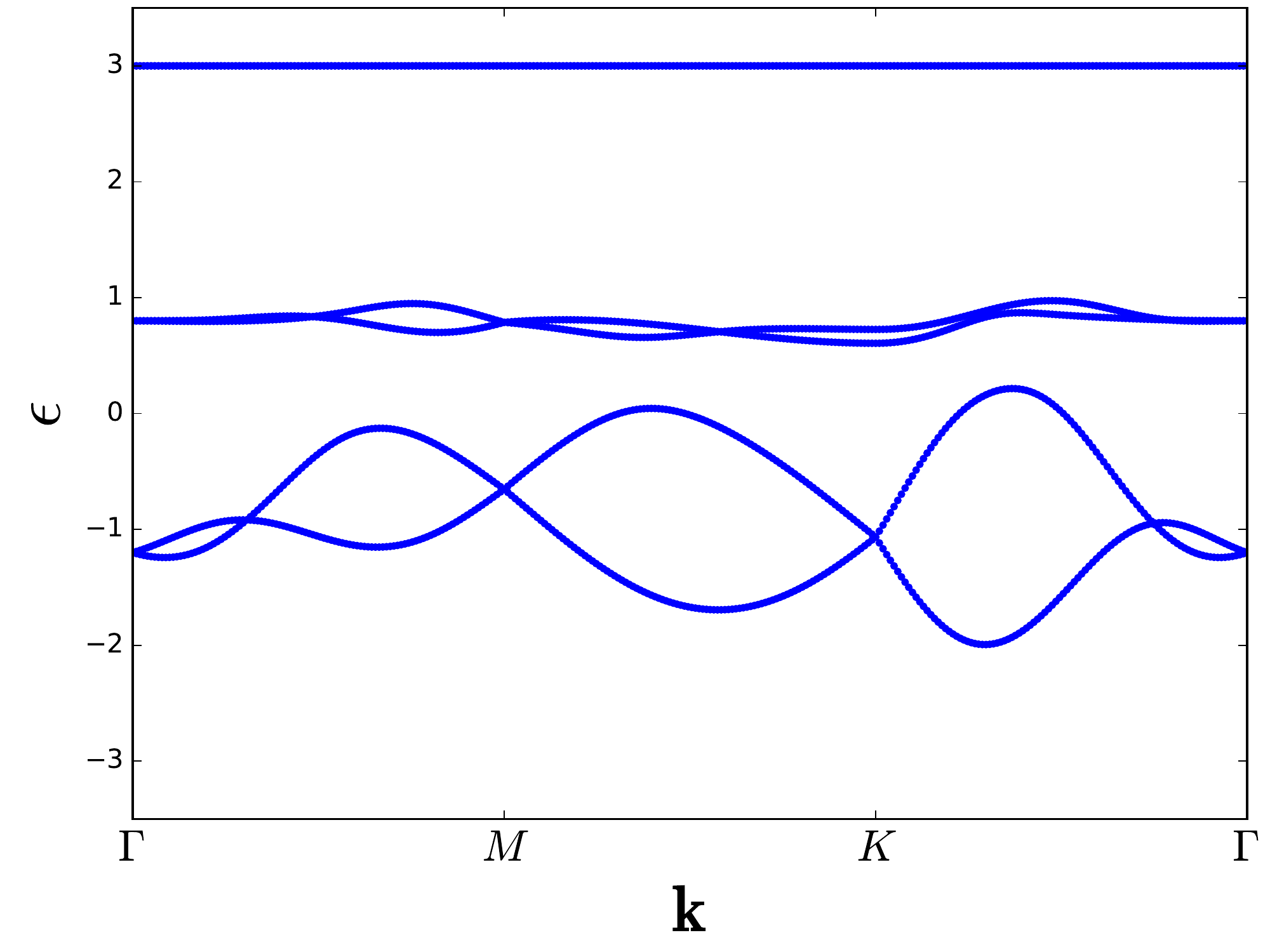}\label{fig:mu0spec}
}
\subfloat[]{
	\includegraphics[width=0.2\textwidth]{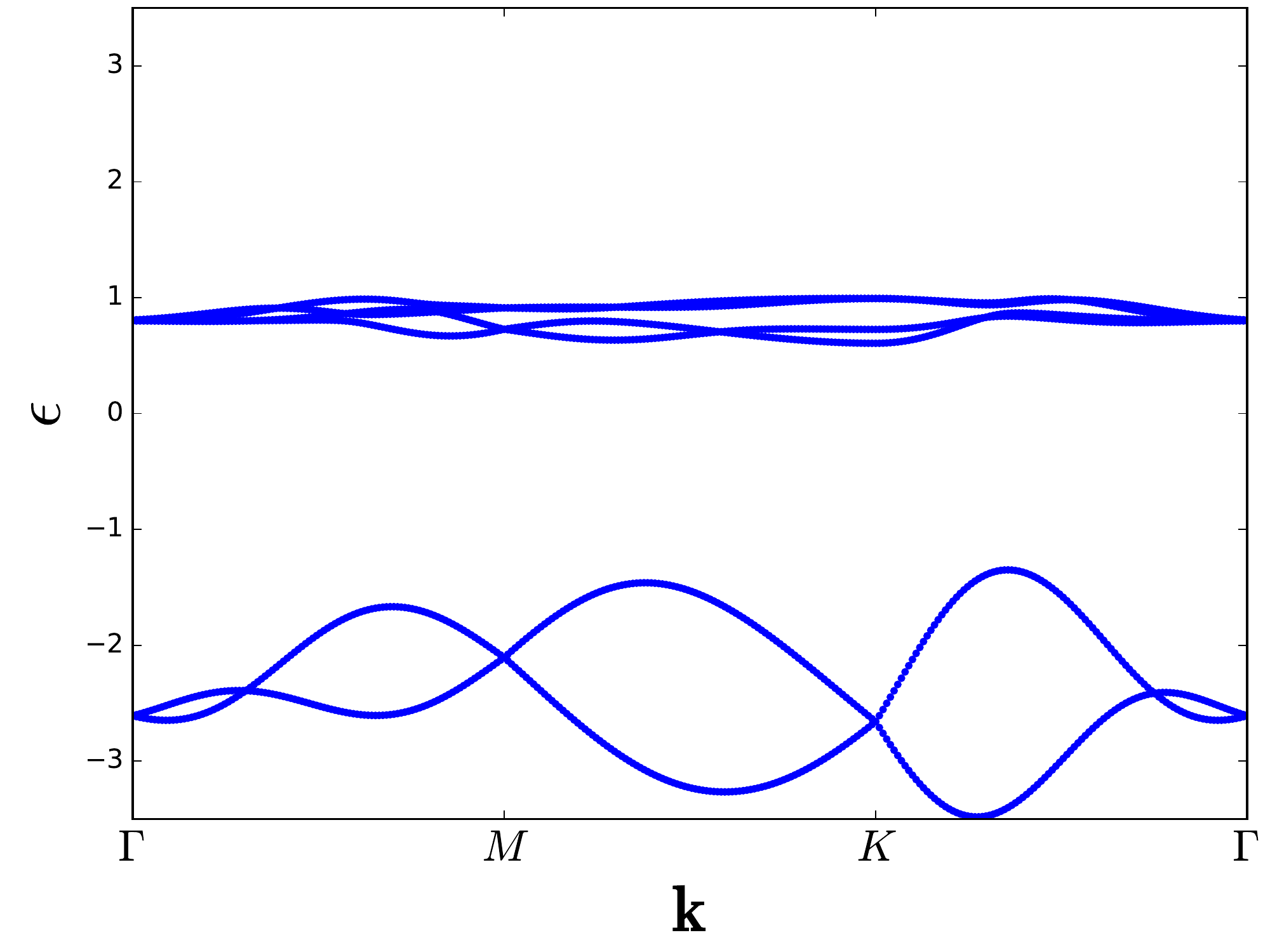}\label{fig:gapclose}
}\qquad
\subfloat[]{
	\includegraphics[width=0.2\textwidth]{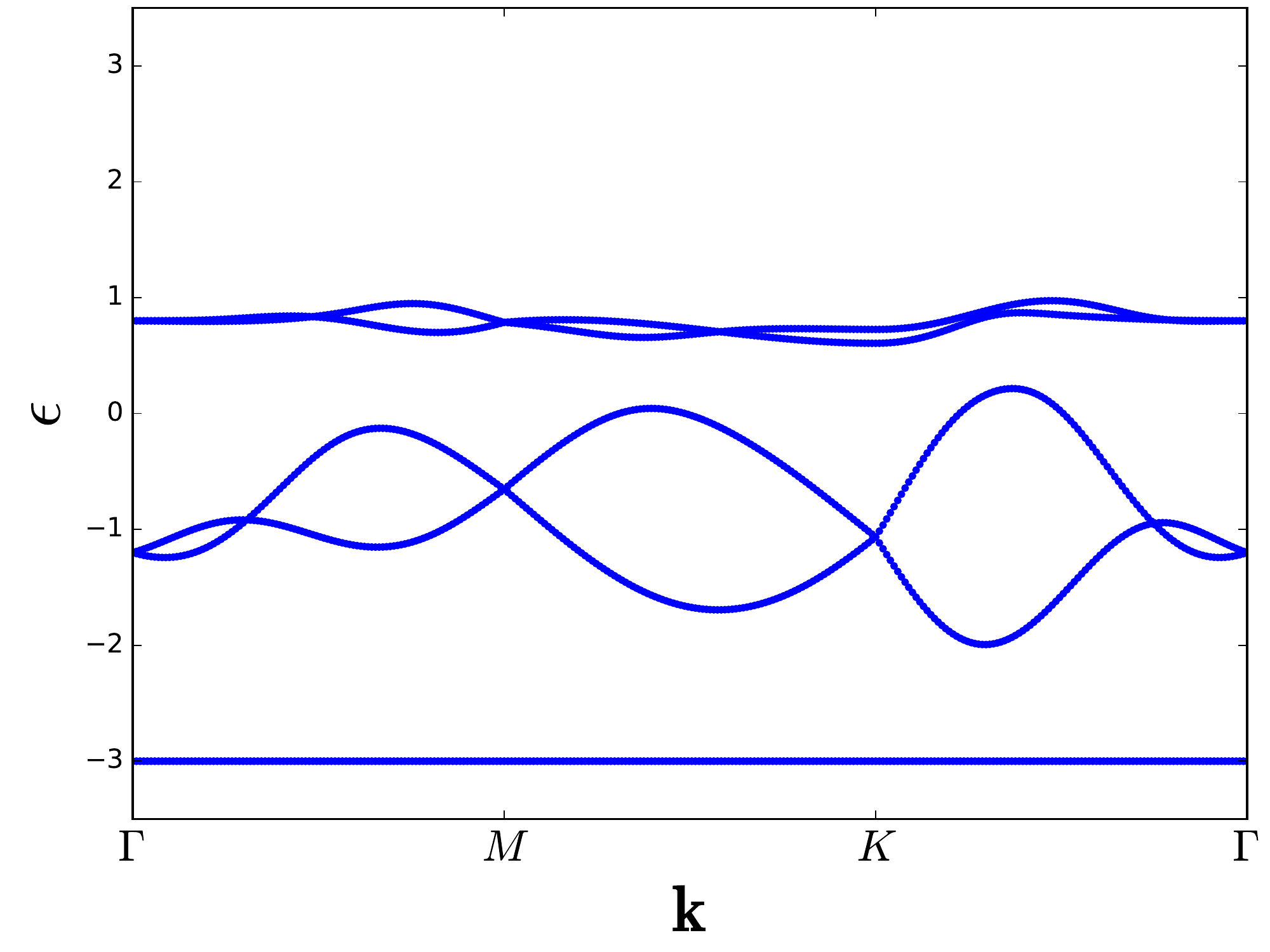}\label{fig:mu1spec}
}
\subfloat[]{
	\includegraphics[width=0.2\textwidth]{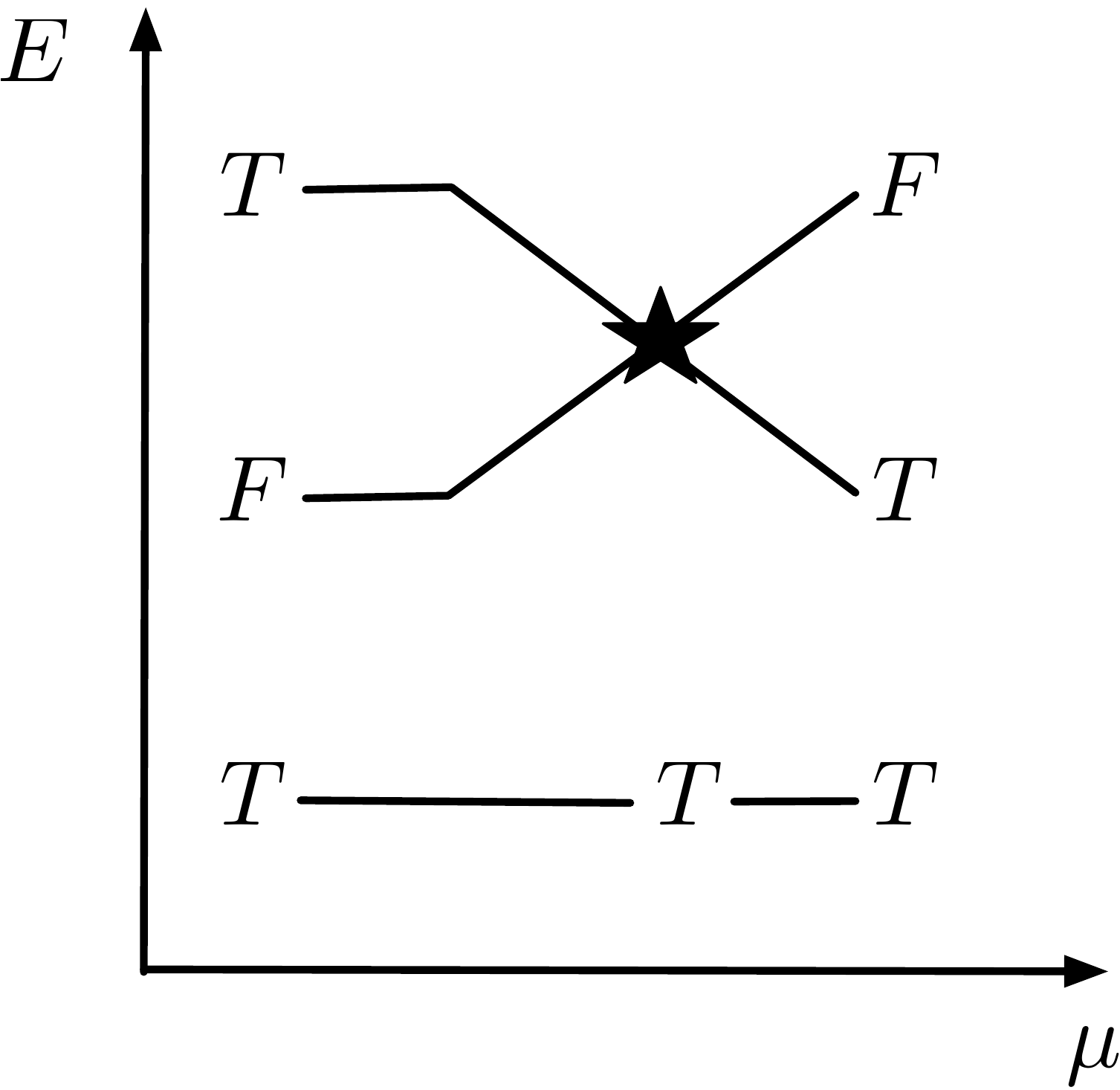}\label{fig:mutune}
}
\caption{Evolution of the spectrum as a function of coupling $\mu$ to ancillary sites. (a) shows the spectrum of the Hamiltonian in Eq.~(\ref{eq:model}) at $\mu=0$, where the ancillary states are decoupled and lie far above the topological bands. (b) shows the model at the point where the gap between the ancillary and topologically nontrivial bands closes, $\mu \approx 0.60$. The upper four bands correspond to the topologically nontrivial bands of the model, overlapping with the ancillary states on the $1a$ position. (c) shows the spectrum at $\mu=1$. The trivial states at the $1a$ position have been teleported to the lowest pair of bands.
The four highest-energy bands in (c) form the same disconnected EBR as the lowest and middle bands in (a). (d) shows schematically the evolution of band topology as $\mu$ is varied from $0$ to $1$. $T$ denotes groups of bands with a non-winding Wilson loop, while $F$ denotes groups of bands with a winding Wilson loop. The black star denotes the region in parameter space in which the four highest energy bands are interconnected.}
\end{figure}

For $\mu=0$, this describes the original four-orbital model, with two additional decoupled bands well separated in energy, as shown in Fig.~\ref{fig:mu0spec}. As $\mu$ is tuned from $0$ to $1$, the lowest pair of bands is deformed into a set of bands originating from atomic-like orbitals centered on the $1a$ site, identical to the ancillary bands at $\mu=0$, as shown in Fig.~\ref{fig:mu1spec}. The authors emphasize that this occurs without closing a gap between the lowest band and the bands above it. However, as a function of $\mu$, a gap \emph{does close} between the (topologically nontrivial) middle pair of bands and the ancillary bands {(a fact which was not explicitly highlighted in Ref.~\onlinecite{comment}). That this must be the case follows from examining the ordering of irreducible representations at the $K$ point at $\mu=0$ and at $\mu=1$: we see from Figs.~\ref{fig:mu0spec} and \ref{fig:mu1spec} that} a band inversion at the $K$ point is required to pass from $\mu=0$, where the highest-energy representation at $K$ is the two-dimensional $\overline{K}_6$ representation, to $\mu=1$, where the representations $\overline{K}_4$ and $\overline{K}_5$ are at a higher energy than $\overline{K}_6$. More generally, this is a consequence of the change of band ordering under the homotopy $\mu$: in order for the disconnected EBR to have the same energy order after $\mu\rightarrow 1$, there must be a band inversion of the trivial $1a$ bands and the topological bands. We show the spectrum of the model at one of the gap-closing points in Fig.~\ref{fig:gapclose}. Throughout this process, there is always at least one group of (two) bands which does \emph{not} admit a localized Wannier description {(except at the gapless point, where a two-band projector is no longer well-defined)}. Once a gap closes and reopens between the topological bands and the ancillary bands, we are forced to re-evaluate which sets of bands transform as a disconnected EBR. We show this process schematically in Fig.~\ref{fig:mutune}.

\begin{figure}[h]
\subfloat[]{
	\includegraphics[width=0.2\textwidth]{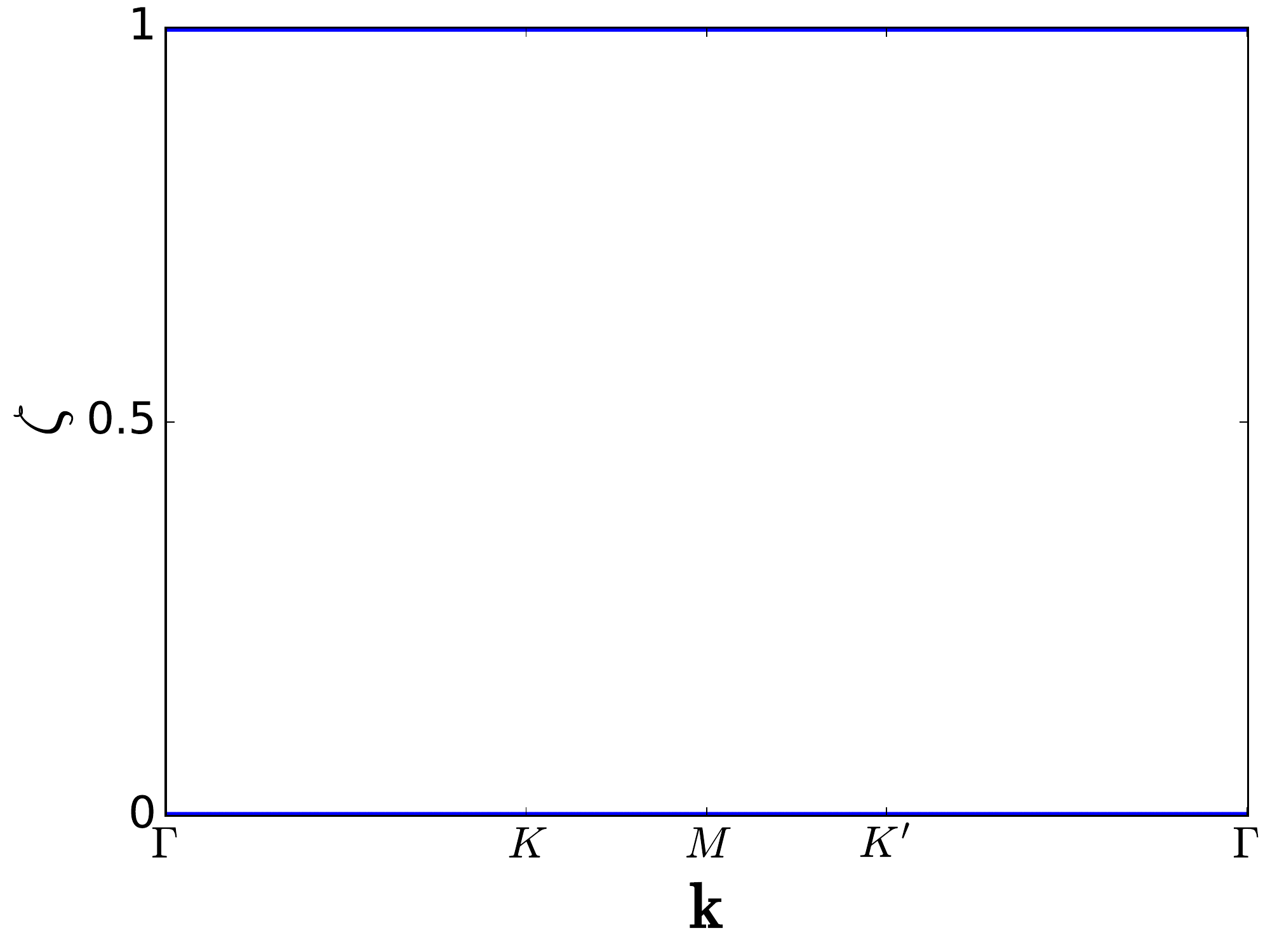}\label{fig:trivialent}
}
\subfloat[]{
	\includegraphics[width=0.2\textwidth]{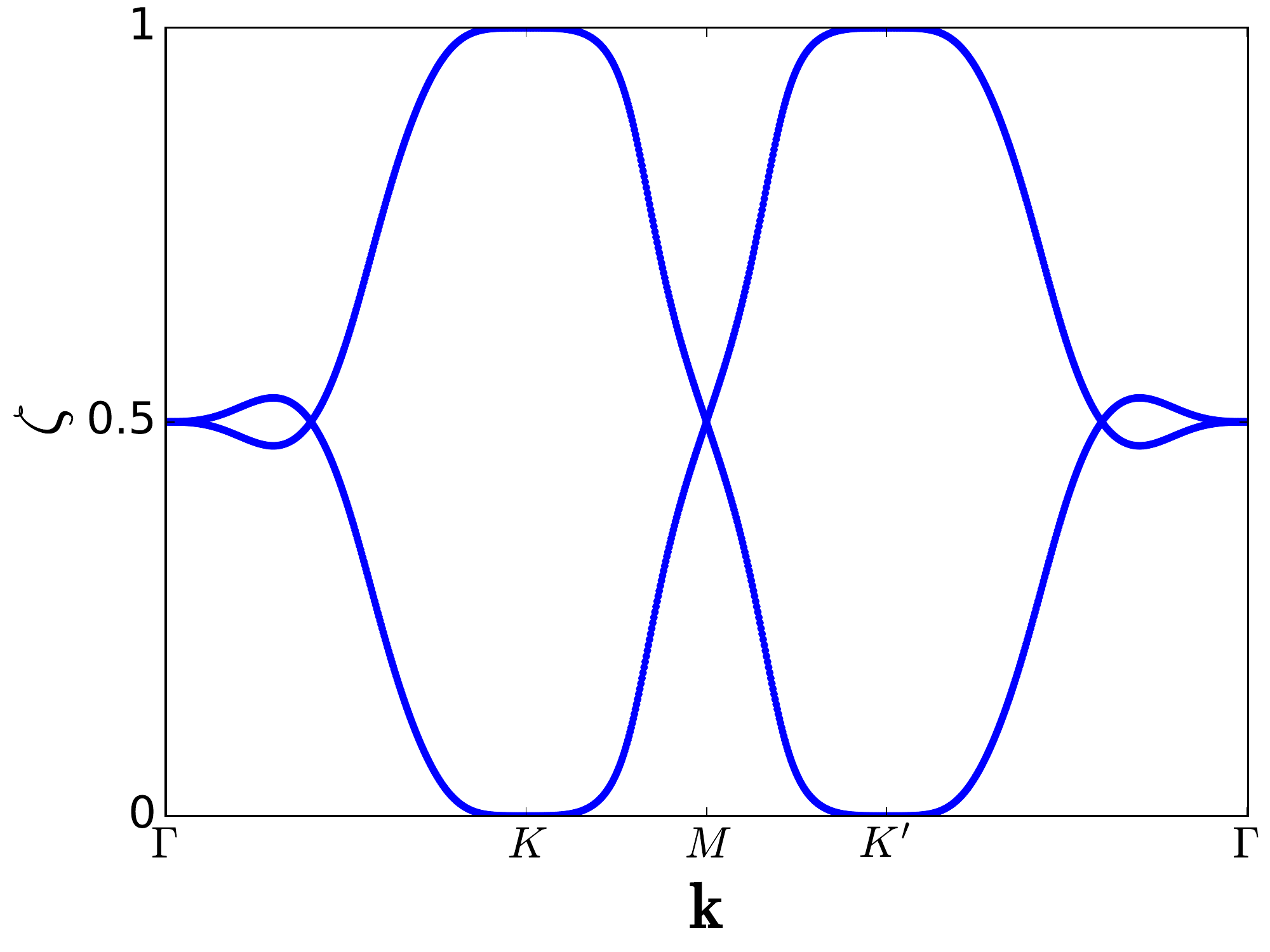}\label{fig:teleporent}
}\qquad
\subfloat[]{
	\includegraphics[width=0.2\textwidth]{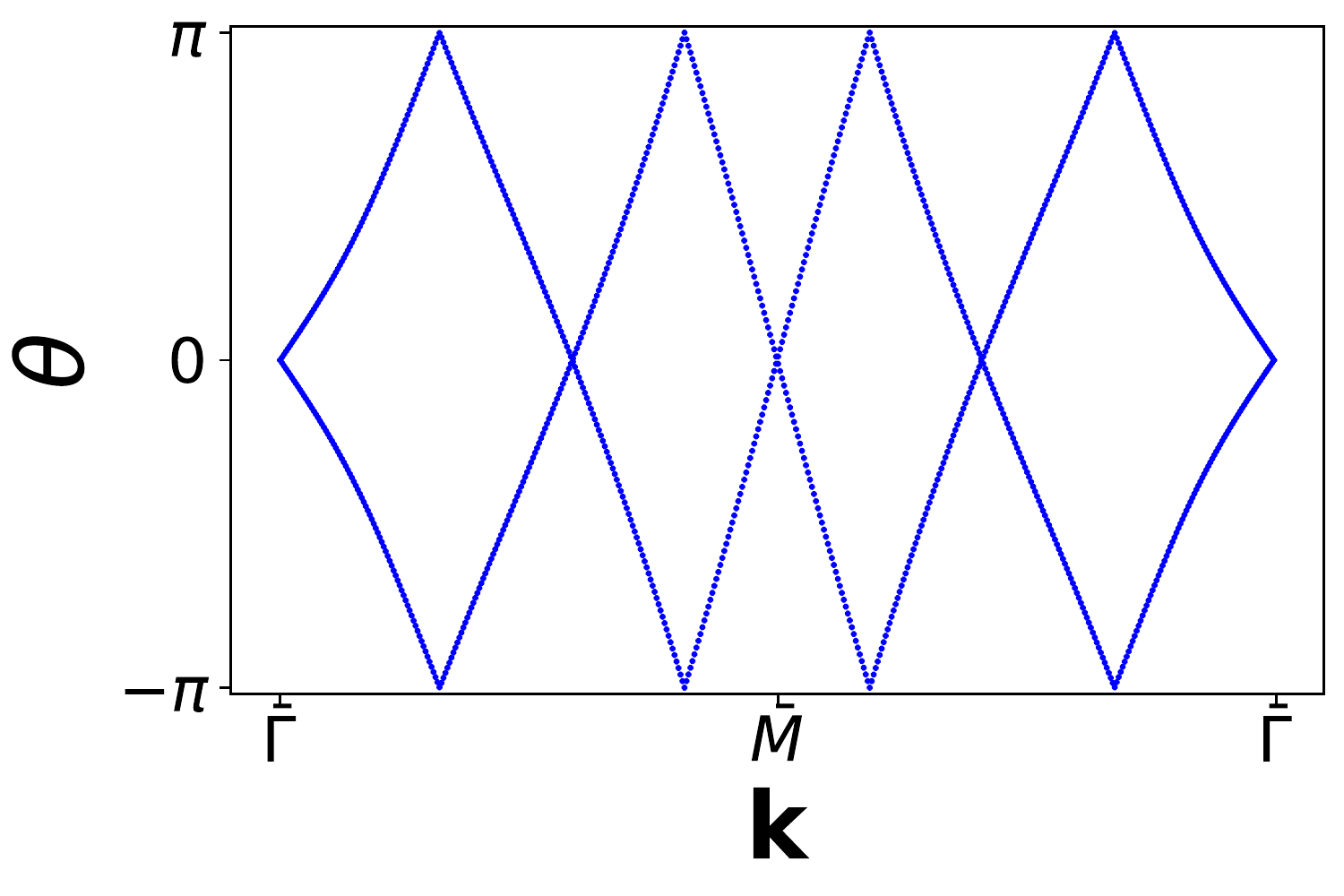}\label{fig:topwilson03}
}
\subfloat[]{
	\includegraphics[width=0.2\textwidth]{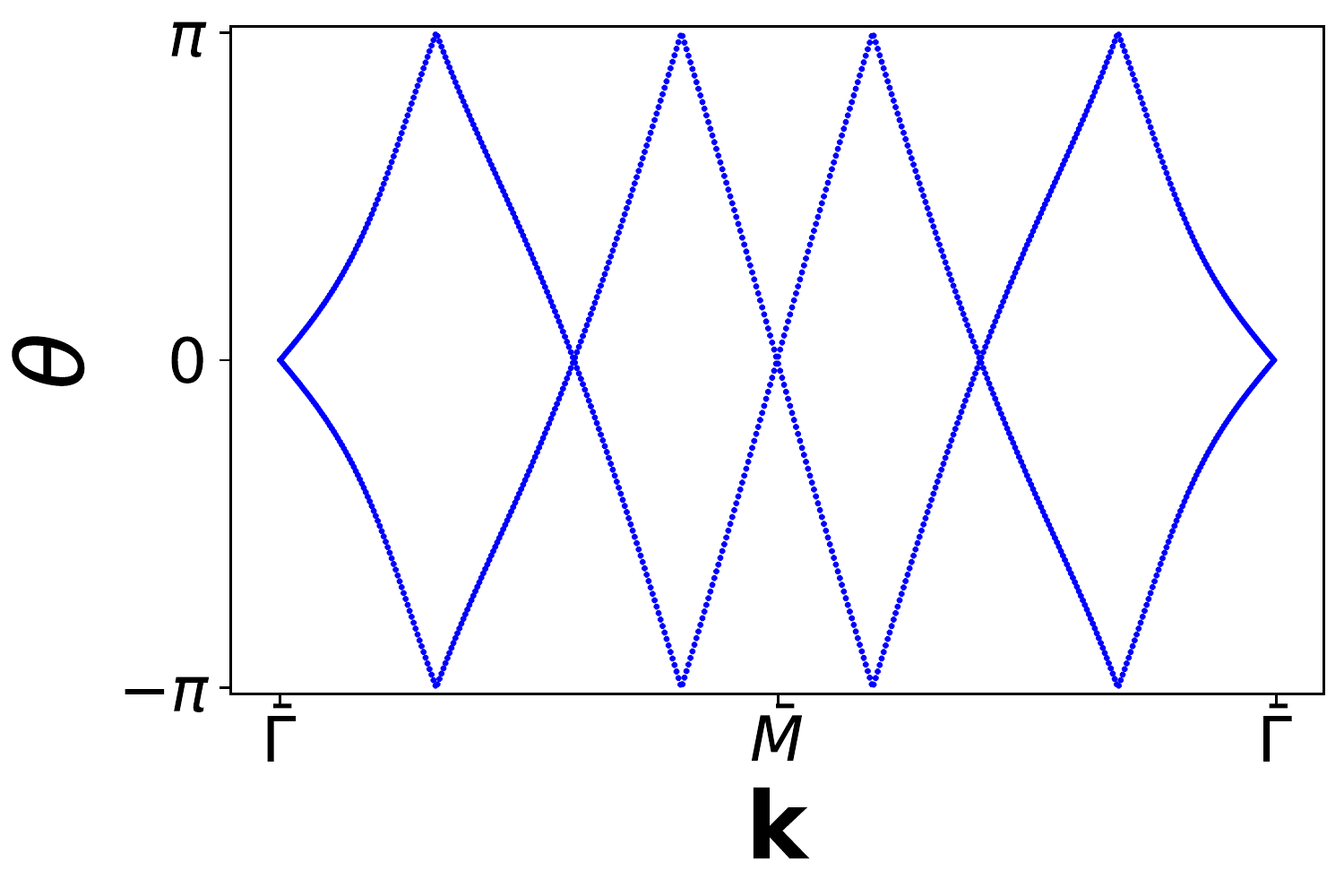}\label{fig:topwilson08}
}
\caption{(a) Translationally-invariant entanglement spectrum for the two lowest bands at $\mu=1$, with an entanglement region containing the $1a$ ancillary site as well as the A sublattice. The entanglement spectrum is trivial: all states have weight $1$ on the $1a$ site, indicating that these bands correspond to atomic-limit orbitals on this site. (b) shows the translationally invariant entanglement spectrum of the two highest bands of the model at $\mu=1$, for the same entanglement cut. Contrary to (a), the entanglement spectrum is nontrivial, matching exactly the spectrum of the two middle bands at $\mu=0$ as shown in Fig.~\ref{fig:0muentspec}. This shows that we have indeed teleported the trivial states from the highest pair of bands at $\mu=0$ to the lowest (at $\mu=1$) pair of bands. (c) Shows the Wilson loop for the topological band at $\mu=0.3$, where it is the middle set of bands in the spectrum. (d) shows the Wilson loop for the same bands at $\mu=0.8$, where it is the highest group of bands in the spectrum. Note that the loops in (c) and (d) are topologically indistinguishable from the loop shown in Fig.~\ref{fig:0muwilsoncond}.}
\end{figure}

Crucially, we have not eliminated any topological bands throughout this process. The Wilson loop evaluated in the subspace of the four highest bands of this model does not wind\cite{Slager2018}; {for $\mu\geq 0.6$ it is the four highest bands that form a four-band EBR, which is connected for the value of $\mu$ depicted in Fig.~\ref{fig:gapclose}. This EBR then splits again for larger values of $\mu$.} There remains for generic values of $\mu$ a set of two topological bands, and hence a suitably chosen two-band Wilson loop which winds, shown in Figs.~\ref{fig:topwilson03} and \ref{fig:topwilson08}. Since the ancillary bands are decoupled from the original bands at $\mu=0$ (Fig.~\ref{fig:mu0spec}) and $\mu=1$ (Fig.~\ref{fig:mu1spec}), in both cases there is a group of bands whose irreps at high-symmetry points are incompatible with a local Wannier description.
Note that we have not trivialized a disconnected EBR.  Rather, this homotopy exhibits a novel ``band teleportation'' effect -- as a function of $\mu$, the trivial states from the 
highest energy band are teleported to the lowest energy band,{without ever closing the gap between the upper four bands and the lower two}. In other words, before a gap between the ancillary bands and the middle two bands closes, it is the lowest four bands which transform as a disconnected elementary band representation; after a gap to the ancillary bands closes and reopens, it is the highest four bands which transform as a disconnected elementary band representation.
To verify this claim, we plot in Figs.~\ref{fig:trivialent} and \ref{fig:teleporent} the entanglement spectrum for the two lowest groups of bands at $\mu=1$. We see that these bands corresponds to trivial atomic orbitals on the $1a$ site, while the middle bands have the same entanglement as the original lower bands of the $\mu=0$ disconnected EBR.  

Such band teleportation is possible precisely due to the nontrivial entanglement present in the topological, disconnected EBR. Closing a gap between the ancillary states and the disconnected EBR
transfers the orbital character of the ancillary states to the (entangled) lower energy band of the EBR. This process is reversible, and ``conserves'' topology in the sense that for all $\mu$ there is always at least one topological set of two bands in the spectrum. To see this, we plot in Figs.~\ref{fig:topwilson03} and \ref{fig:topwilson08} the Wilson loop of the topological bands for $\mu=0.3$ and $\mu=0.8$ respectively. We see that in both cases, the Wilson loop spectrum is topologically indistingushable from the $\mu=0$ result of Fig.~\ref{fig:0muwilsoncond}.
Finally, if the ancillary bands are removed from the spectrum, the resulting Hamiltonian is identical to the Hamiltonian before the ancillary bands were included, leaving behind exactly the starting point. We note that similar pheomena may be observable in bands forming an obstructed atomic limit.

The refined notion of topology of projectors onto isolated groups of bands, rather than of the stable topology of valence bands, is relevant for finding new topological material candidates, where any isolated group of bands may be experimentally accessed, even in the ``unoccupied'' bands, by pumping experiments. {TQC can capture both the ``strong'' (stable) and the fragile instances of band topology.}

We see then that this model for a ``fragile'' topological phase corresponds to one of the disconnected EBRs given in the exhaustive classification of Refs.~\onlinecite{NaturePaper,GroupTheoryPaper} of all 10398 EBRs and PEBRs in the non-magnetic space groups. Recall there that an EBR was defined not just in terms of the little group representation of bands at high-symmetry points, but in terms of the smooth dependence of the representation matrices (sewing matrices) as a function of $\mathbf{k}$ for all momenta in the Brillouin Zone (note that with this definition, distinct EBRs may have the same irreps at high symmetry points, but are distinguishable nevertheless through Wilson loop or sewing matrix invariants\cite{}). Only one of two groups of bands that together comprise the disconnected EBR admits exponentially localized Wannier functions, but the other \emph{cannot}.
A distinct set of  bands in these phases can be continued to a non-obstructed atomic limit if ancillary bands are added to the Hilbert space, reminiscent of the spinless topological phases of Ref.~\onlinecite{Alexandradinata14}. However, this comes at the price of having the band gap close between {a pair of ancillary bands and the topological bands}. We call this process ``band teleportation".
However, at all times in the process, there exist topological bands, i.e. bands not describable by localized Wannier orbitals. These phases fall outside of the classification schemes predicated on stable equivalence, which are, by definition, robust under adding arbitrary trivial bands. Thus, although the little group representations that appear in this model are sufficient to determine that the bands cannot be topologically trivial, it is not included in the classification of Ref.~\onlinecite{Po2017}. The EBR theory (and, more generally, TQC) gives a more complete description of topological electronic bands. The most general  experimentally relevant theory defines a set of topological bands as one which does not transform as band representations (of which a disconnected elementary band representation is the simplest example).

We emphasize again that the non-trivial two-band Wilson loops shown in Figs.~\ref{fig:0muwilsoncond},\ref{fig:topwilson03}, and \ref{fig:topwilson08} constitute a robust and physically meaningful signature of the nontrivial topology in this model. In particular, the topological winding of a two-band Wilson loop cannot be changed without closing a gap between the topological bands and other bands in the spectrum, due to the stability of the winding proved in Appendix~\ref{app:c2tproof}. Furthermore, even though generic Wilson loops with $N$ total bands in this space group will not wind, a two-band Wilson loop can always be defined for a pair of bands separated by a spectral gap from all other states in the spectrum. Thus, we explicitly show that the models here consisting of a disconnected EBR plus ancillary bands indeed do contain topologically nontrivial bands, distinguished by an in-principle measurable invariant (the Wilson loop winding). We see then that the Wilson loop ``in the space of all occupied bands,'' is insufficient to characterize the topology of a material; {more involved Wilson loops must be considered.}

\section{Topological bands and Wilson Loop Windings}\label{sec:notrhex}

\begin{figure}
\subfloat[]{
	\includegraphics[width=0.2\textwidth]{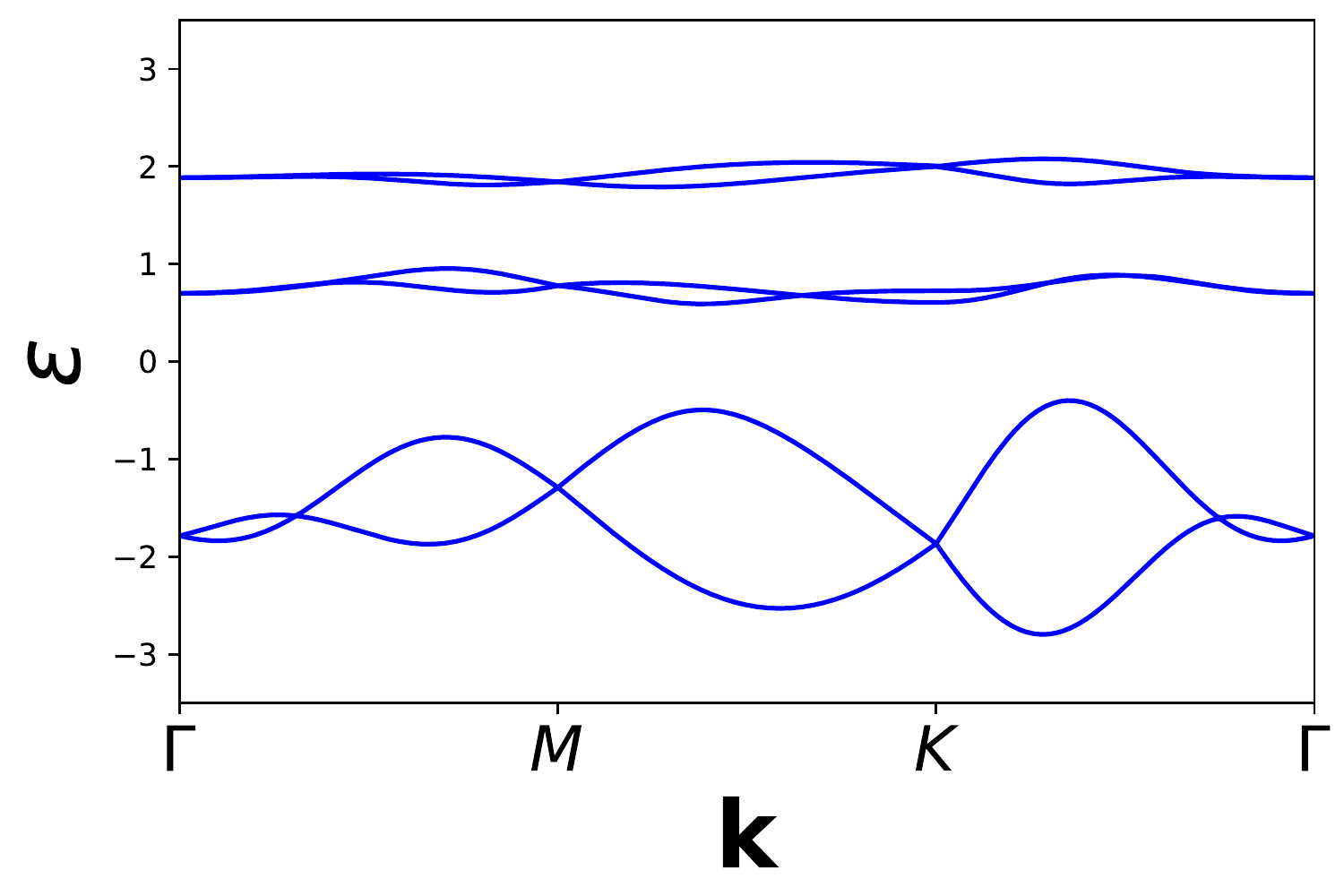}\label{fig:p6mmspec}
}
\subfloat[]{
	\includegraphics[width=0.2\textwidth]{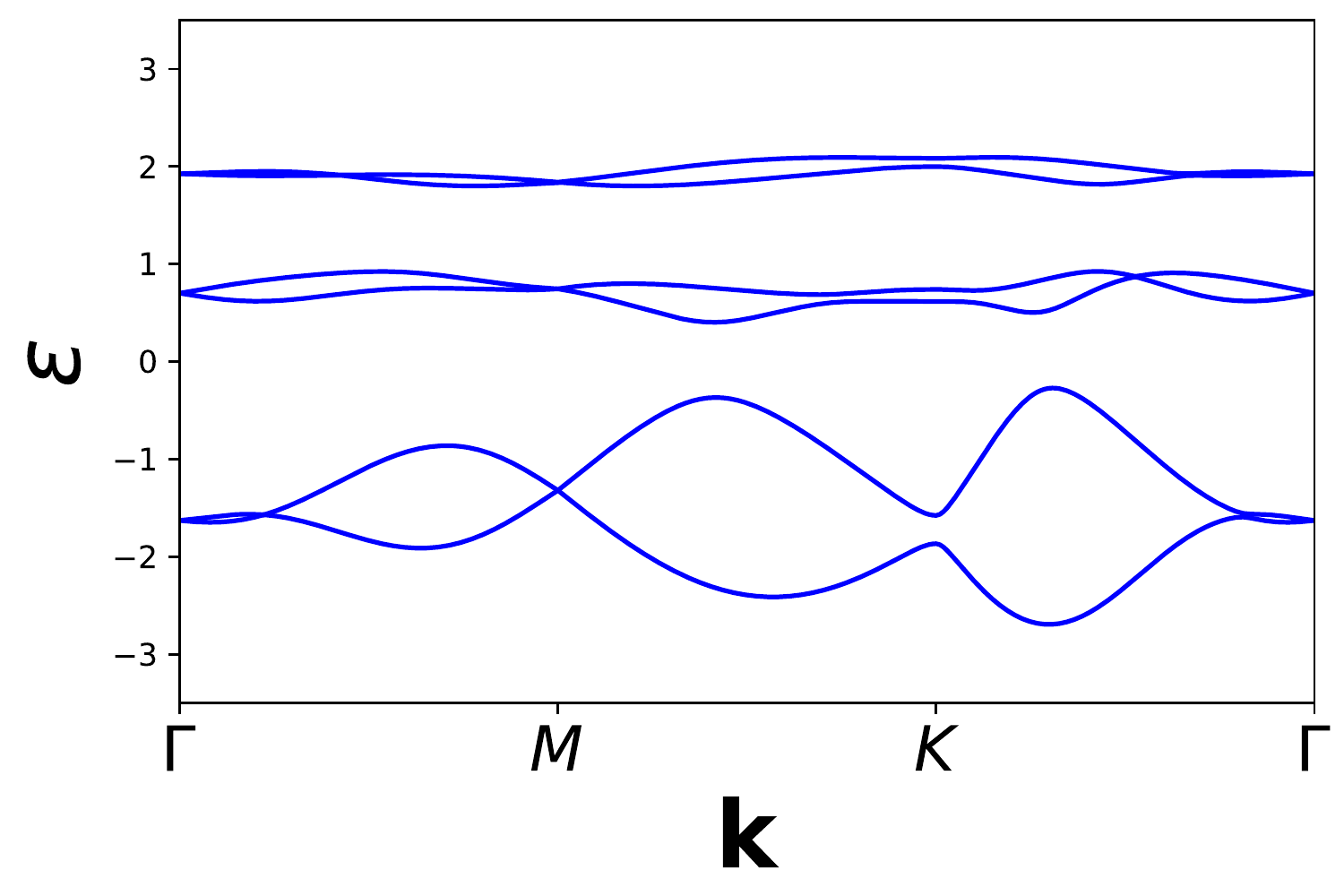}\label{fig:p3m1spec}
}
\caption{Spectra for six band models obtained by breaking symmetries in the original model of Sec.~\ref{sec:model}. (a) is obtained by breaking time reversal symmetry while preserving all spatial symmetries of $p6mm$. (b) is obtained by then further breaking $m_x$ symmetry, leaving a model invariant under the space group $p3m1$. In both cases, the middle group of bands fails to admit a description in terms of localized Wannier functions.}
\end{figure}

From the preceding discussion, we see that the topological nontriviality of bands in this model is reflected in the $C_2T$-protected winding of the Wilson loops directed along the reciprocal lattice vectors. However, it is important to note that, as we mentioned in Sec.~\ref{sec:model} above, the nontriviality of these bands is reflected directly in terms of their little group irreps at $\Gamma$ and $K$. In fact, $C_3$ eigenvalues alone are sufficient to determine that these bands cannot be continued to an atomic limit. In particular, if we break time-reversal symmetry in space group $p6mm1'$ to yield space group $p6mm$ (183), these bands still originate from the disconnected EBR $(\bar{E}\uparrow G)_{2b}$. Going even further, we can break $C_2$ symmetry 
as well while preserving the mirror symmetry $m_y$ which exchanges $K$ and $K'$; doing so reduces the model to space group $p3m1$ (156) (still without time reversal symmetry). {This splits the $2b$ Wyckoff position into the $1a$ position consisting of the $A$ honeycomb sublattice, and the $1b$ position consisting of the $B$ honeycomb sublattice.} In this space group, these two bands are still topologically nontrivial, being expressible as the difference
 $(\bar{E}_1\uparrow G)_{1b}\oplus (\bar{E}_1\uparrow G)_{1b} \ominus (\bar{E}_1\uparrow G)_{1a}$ 
 of EBRs. The symbol ``$\ominus$'' denotes the difference of EBRs as defined in Refs.~\onlinecite{Fragile2017} and \onlinecite{Po2017}. In Appendix~\ref{app:models}, we give the modifications to the parameters of model necessary to realize these space groups. For the remainder of this section, we will take $\mu=0.3$ for all models. In Figs \ref{fig:p6mmspec} and \ref{fig:p3m1spec} we show the spectra for each of these models.

\begin{figure}
\subfloat[]{
	\includegraphics[width=0.2\textwidth]{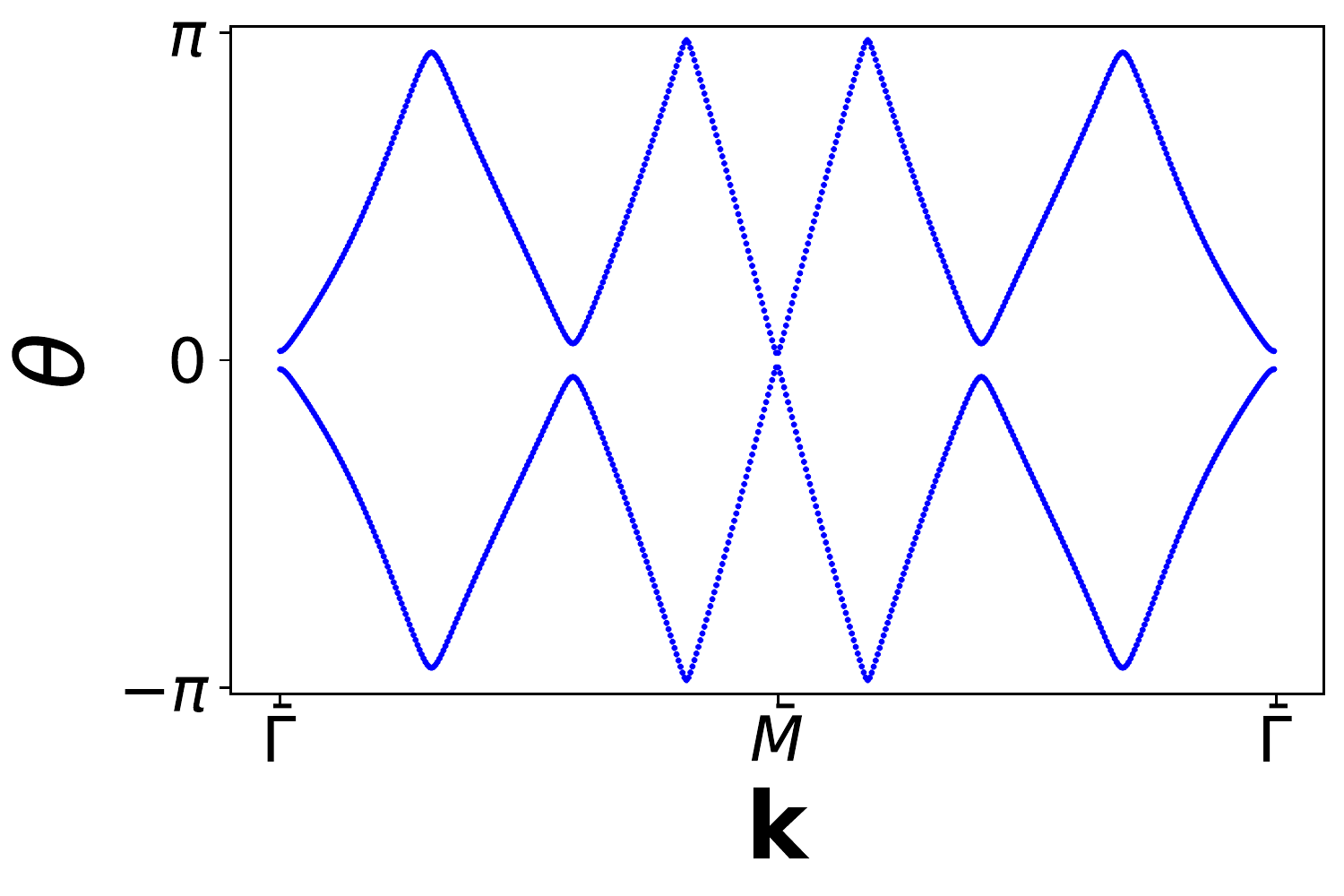}
}\qquad
\subfloat[]{
	\includegraphics[width=0.2\textwidth]{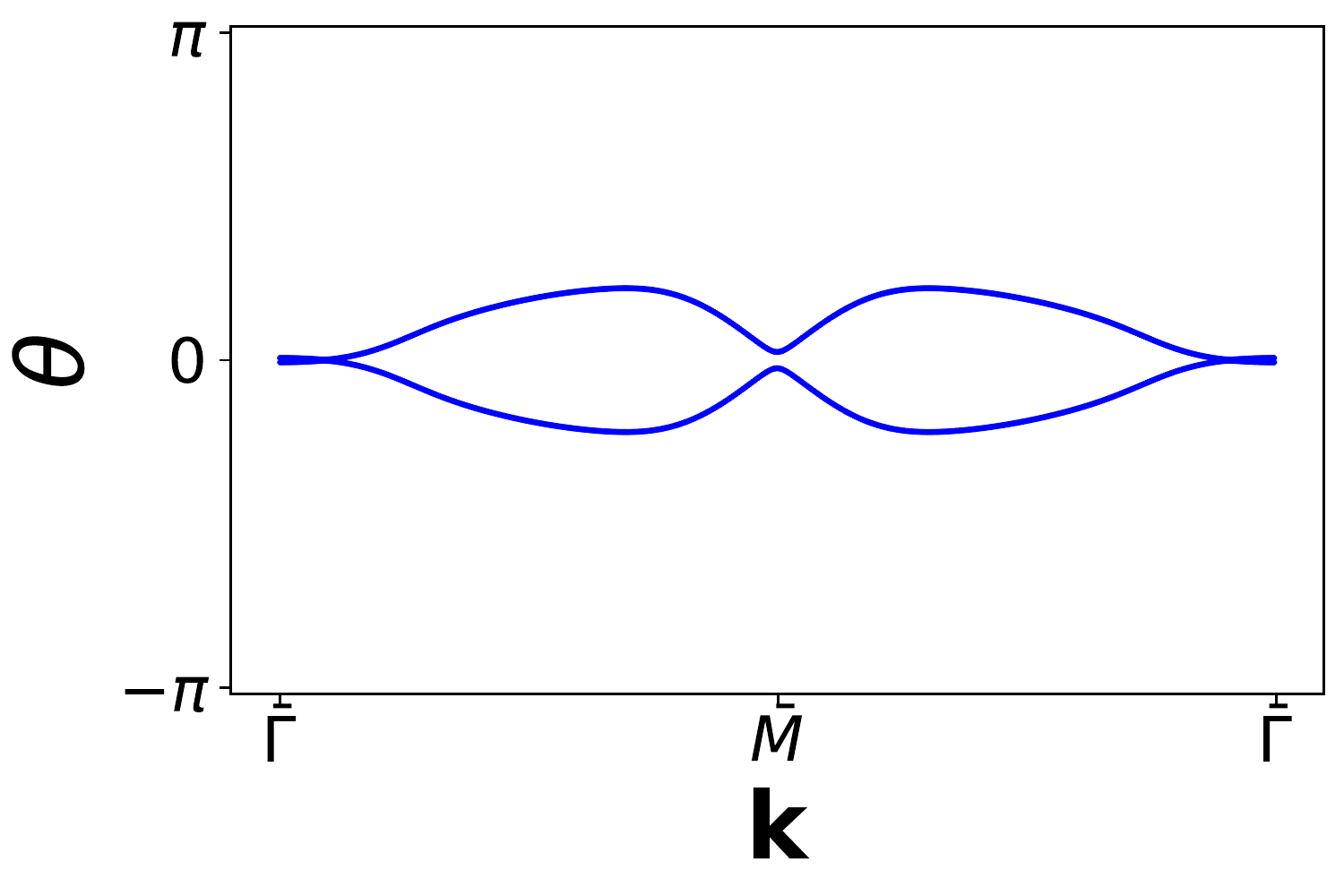}
}\qquad
\subfloat[]{
	\includegraphics[width=0.2\textwidth]{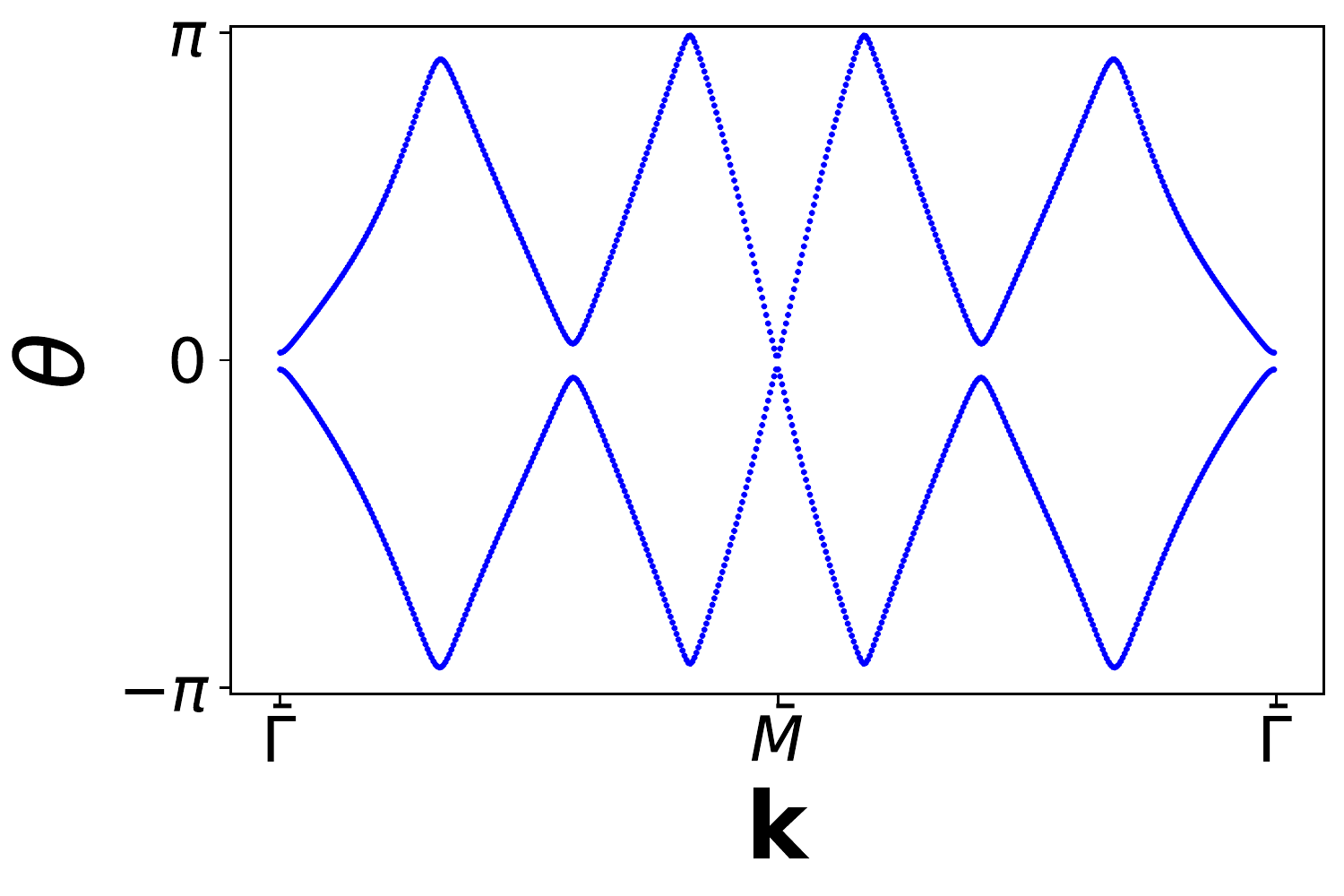}
}\qquad
\subfloat[]{
	\includegraphics[width=0.2\textwidth]{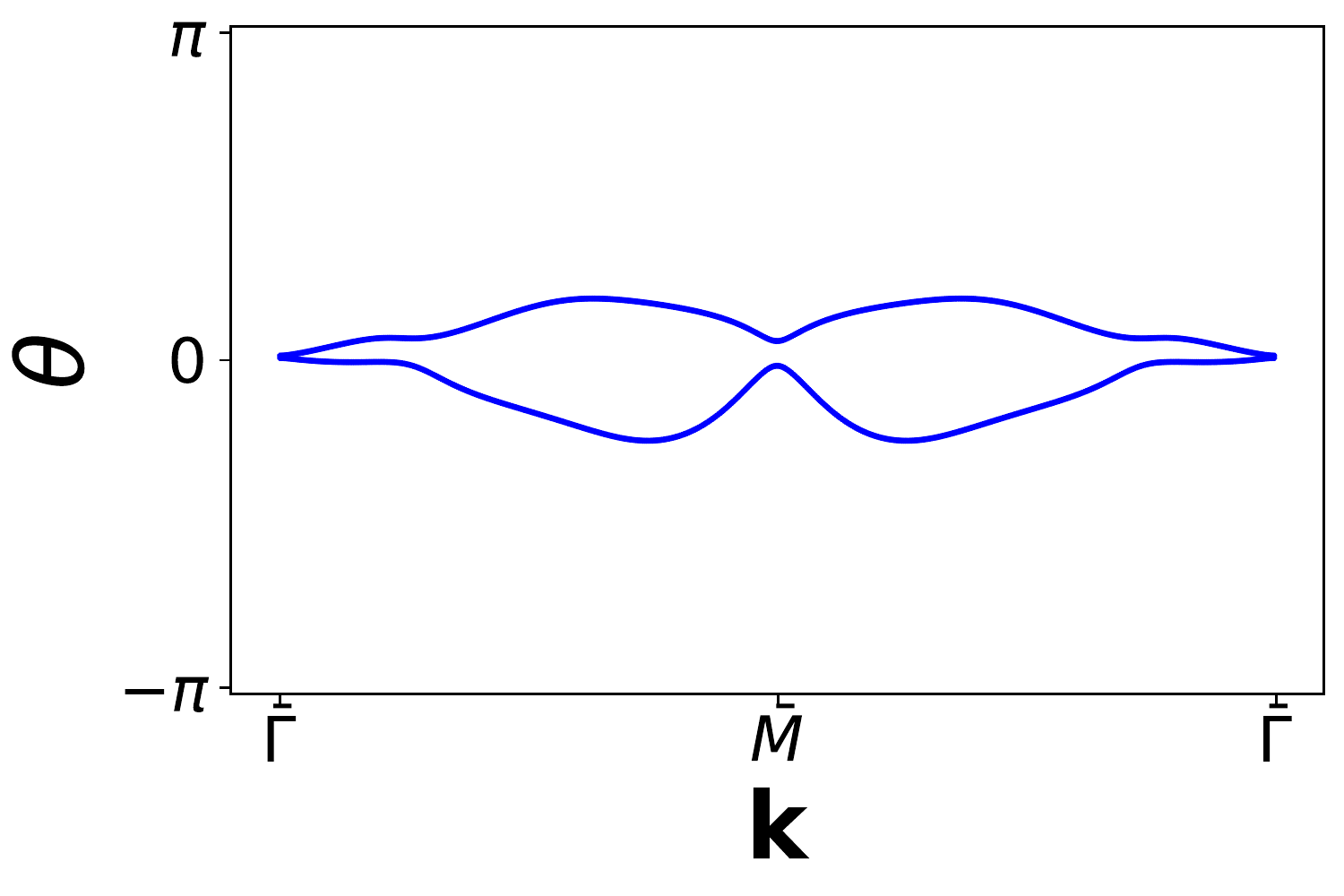}
}
\caption{
Wilson loops for the models in space groups $p6mm$ and $p3m1$.
 (a) shows the Wilson loop for the topological bands in $p6mm$; clear gaps near $\theta=0$ and $\theta=\pi$ can be observed, showing that the loop is deformable to a trivial Wilson loop. 
 (b) shows the Wilson loop for the lowest bands in $p6mm$, which is also clearly does not wind. Similarly, 
 (c) shows the Wilson loop for the topological bands in $p3m1$, and 
 (d) shows the loop for the lowest bands in $p3m1$.}\label{fig:nowind}
\end{figure}

Because we have broken the $C_2T$ symmetry which protects crossings in the Wilson loop discused in Section~\ref{sec:model}, the $\mathbf{g}_1$-directed Wilson loops fail to wind in all of these models. We show this in Fig.\ref{fig:nowind}. This may 
lead one to conclude that all bands in this model are trivial; as we mentioned previously,  this would be incorrect, as can be \emph{easily} checked from little group representations. However, it is natural to ask if there exists a more exotic Wilson loop which reflects the nontriviality of the middle bands in these models. {This reflects the general need to work hard in finding the appropriate Wilson loop needed to show the topological nature of a set of bands. This is especially true in cases where sets of larger numbers of bands can be shown by symmetry eigenvalues to not admit localized Wannier functions.}

\begin{figure}
\subfloat[]{
\includegraphics[width=0.2\textwidth]{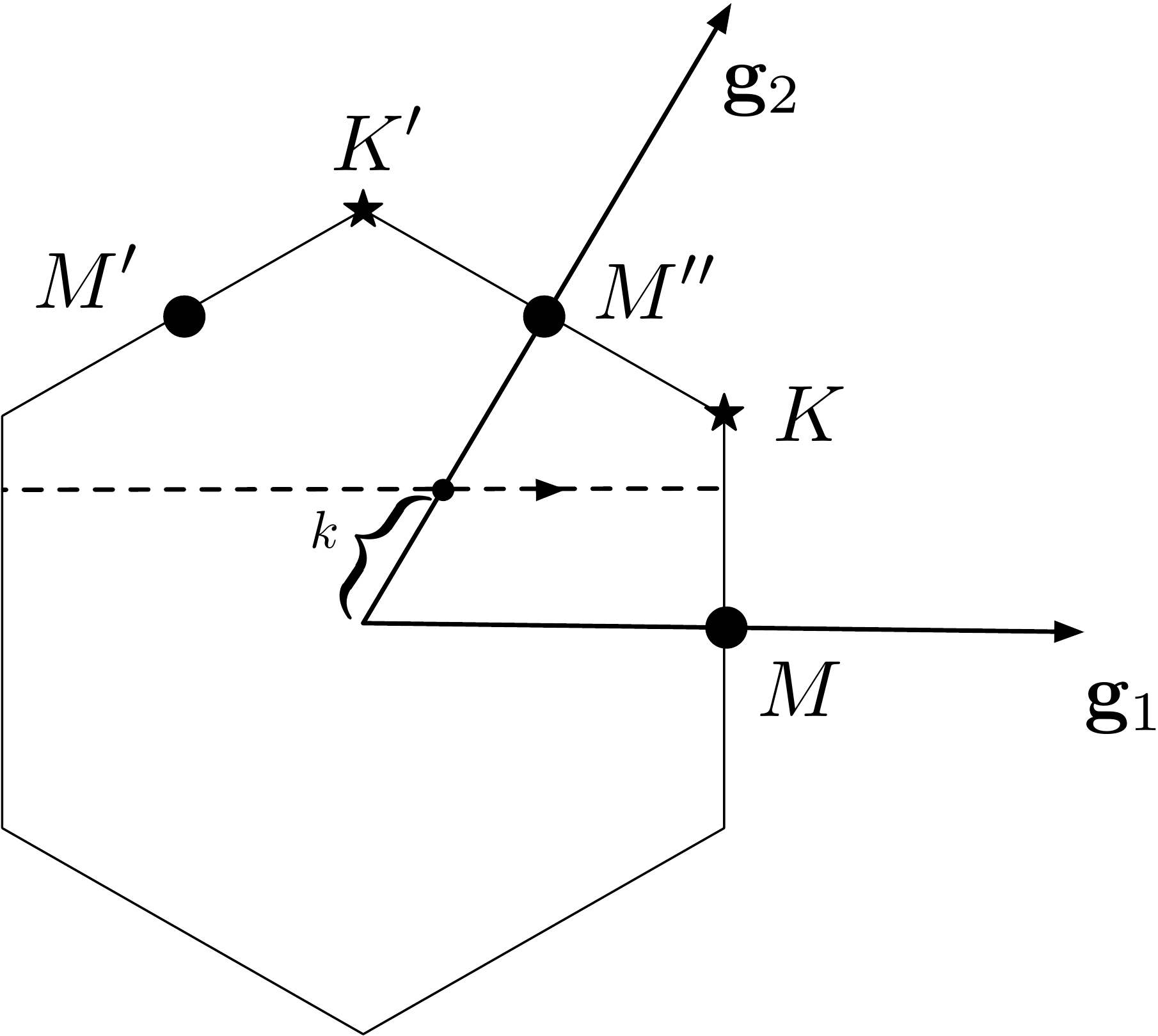}\label{figstrtloop}	
}
\subfloat[]{
\includegraphics[width=0.2\textwidth]{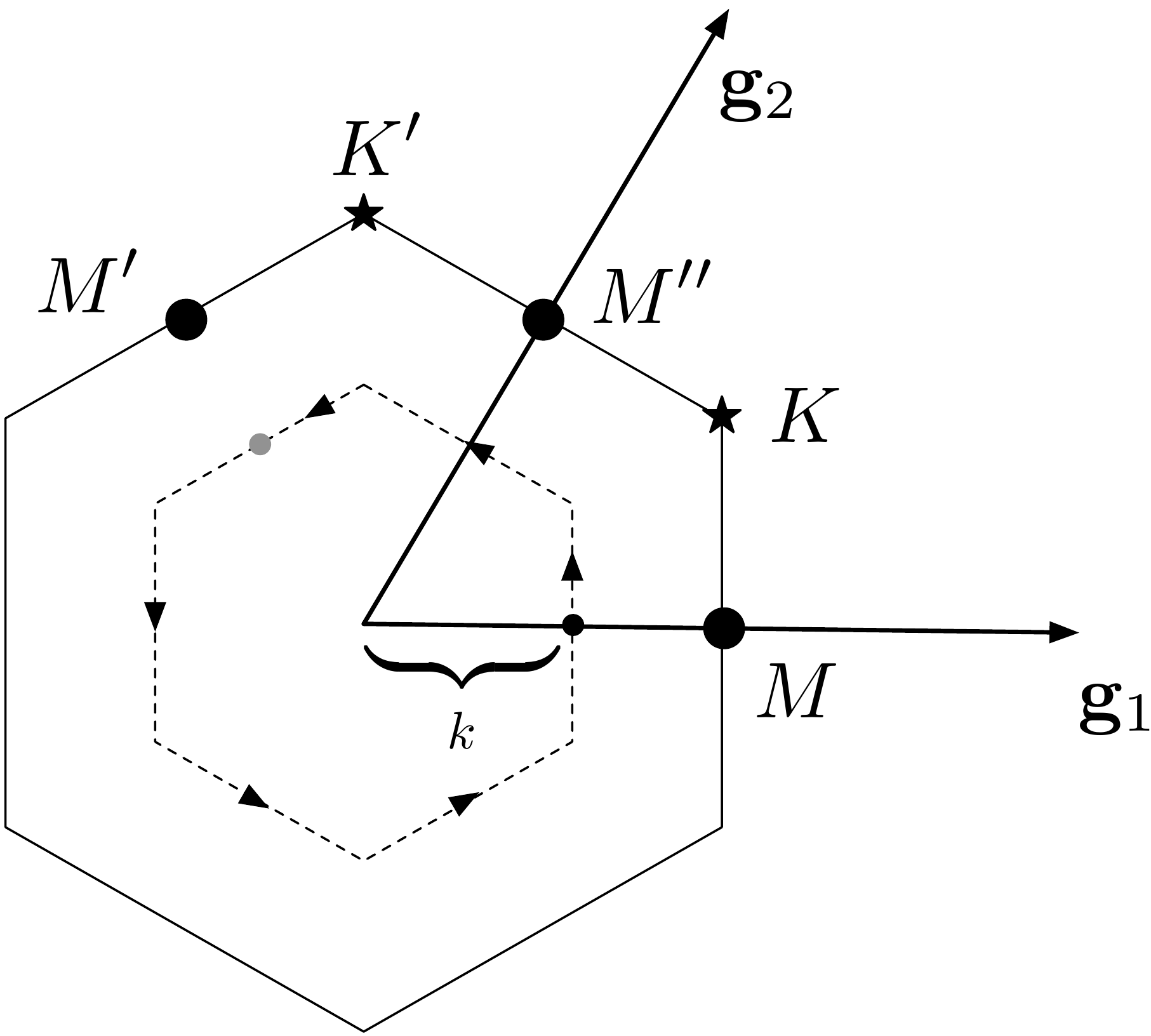}\label{fighexloopdiag}
}
\caption{Diagram of the Brillouin zone in space groups $p6mm$ and $p3m1$, showing the Wilson loop paths used in this work. The reciprocal lattice vectors and high symmetry points are also shown. (a) Path of the straight Wilson loop $W_{\mathbf{g}_1}(k)$ indicated as a dotted line. The basepoint of the loop is shown as a black circle. (b) Path of the hexagonal Wilson loop $W_h(k)$ as a dotted hexagon, with direction indicated; the basepoint of the loop is indicated with a black circle. The endpoint of the Wilson line in the definition of $W_3(k)$ is shown as a grey circle.}
\end{figure}

The importance of $C_3$ symmetry suggests we examine a loop which explicitly respects this symmetry. As such, let us examine the set of hexagonal Wilson loops $W_h(k)$ depicted in Fig.~\ref{fighexloopdiag}. We can express this loop as the cube of a ``dressed'' (i.e.~augmented by multiplication by a symmetry operation) Wilson line\cite{ArisCohomology}
\begin{align}
W_h(k)&=(W_3)^3 \\
W_3(k)&=C_3^{-1}W_{(-k,k)\leftarrow (-2k/3,4k/3)}\times\nonumber \\
&\times W_{(-2k/3,4k/3)\leftarrow(2k/3,2k/3)}W_{(2k/3,2k/3)\leftarrow(k,0)},
\end{align}
Where the Wilson line 
\begin{equation}
W_{\mathbf{k}_2\leftarrow\mathbf{k}_1}=\prod_{\mathbf{k}}^{\mathbf{k}_2\leftarrow\mathbf{k}_1}P(\mathbf{k})
\end{equation}
is an operator in the space of Bloch functions given as a path-ordered product of projectors $P(\mathbf{k})$ onto a suitably chosen set of bands, and $C_3^{-1}$ is the operator effecting an inverse threefold rotation on the space of Bloch functions at the endpoint of the Wilson line.

\begin{figure*}
\subfloat[]{
	\includegraphics[width=0.4\textwidth]{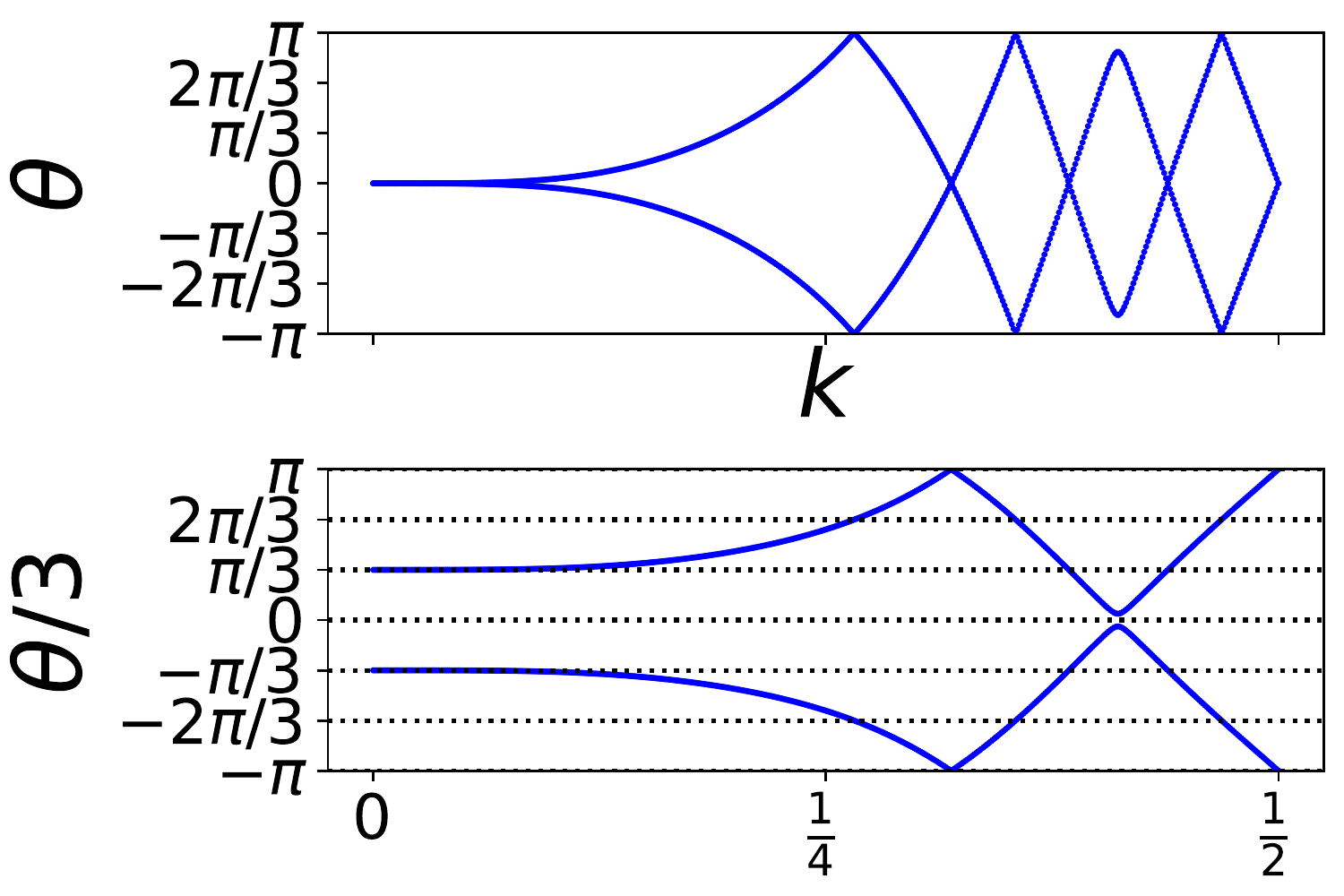}
}
\subfloat[]{
	\includegraphics[width=0.4\textwidth]{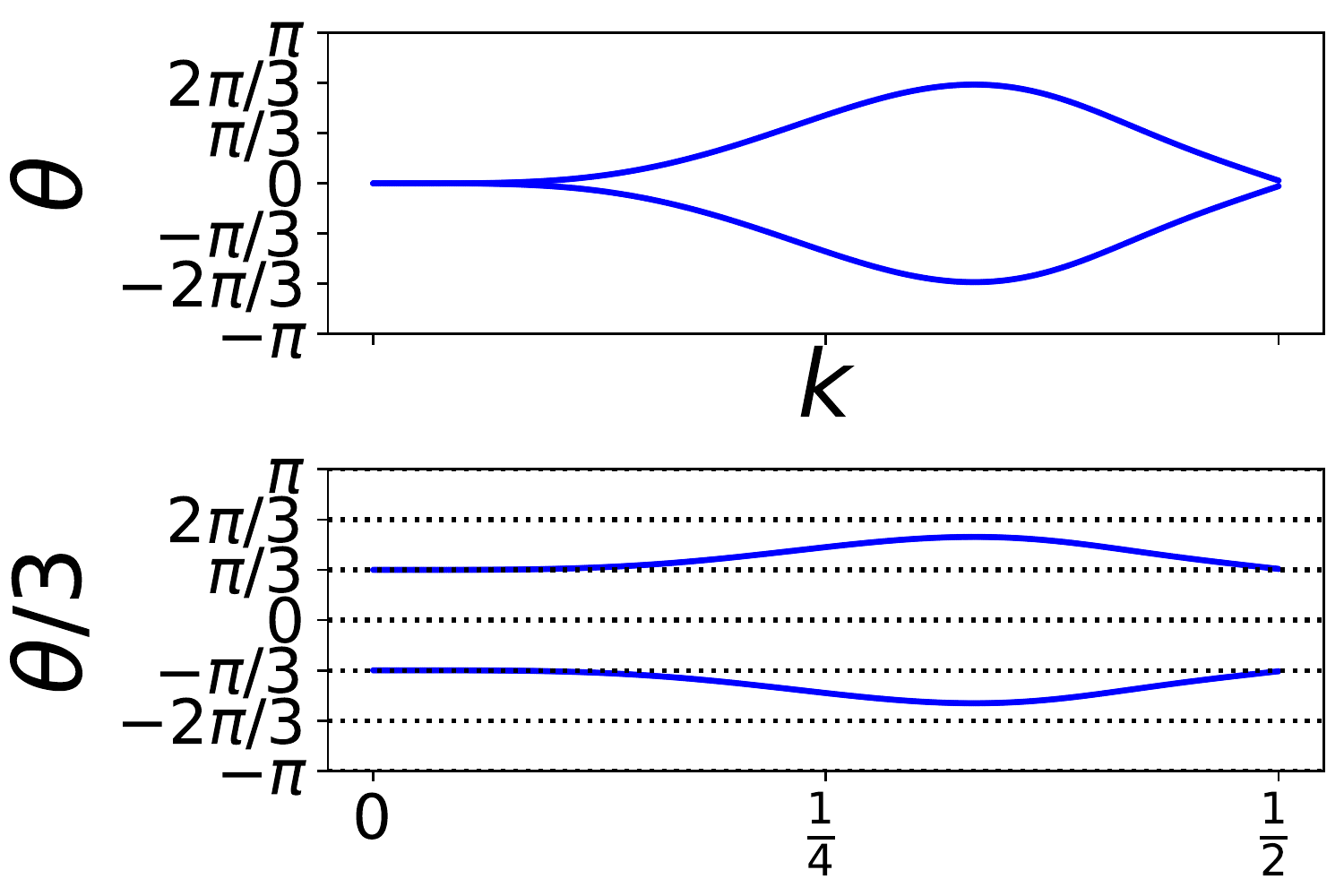}
}
\caption{Spectra of $W_h(k)$ and $W_3(k)$ for each group of two bands in the lowest four bands in our model in space group $p6mm$ without time-reversal symmetry. {Both spectra are pinned at the endpoints $k=0$ and $k=1/2$ by $C_3$ symmetry, as shown in Appendix~\ref{app:hexloopproofs}. The upper plots are obtained from those below by multiplication by three ($\mod 2\pi$), leading to crossings in the spectrum of $W_h$.} (a) shows the spectra for the middle, topological bands. The upper panel is the spectrum of $-i\log W_h(k)$, while the lower panel is the spectrum of $-i\log W_3$(k). We show in dotted lines in the lower panel the values $\theta=n\pi/3$ for integer $n$; {note the avoided crossing near $\theta/3=0$ (the crossing at $\theta/3=\pi$ near $k=1/4$ is protected by $C_6$ symmetry, as explained in Appendix~\ref{app:hexloopproofs})}. Upon taking the cube of this spectrum, these eigenvalues correspond to crossings in the spectrum of $W_h(k)$. The spectrum of $W_h(k)$ exhibits a nontrivial winding, reflecting the topology of these two bands. (b) Shows the same spectra for the lowest two bands. Here we see that there is no winding of the eigenvalues of $W_h(k)$.}\label{fig:hexloop6mm}
\end{figure*}

\begin{figure*}
\subfloat[]{
	\includegraphics[width=0.4\textwidth]{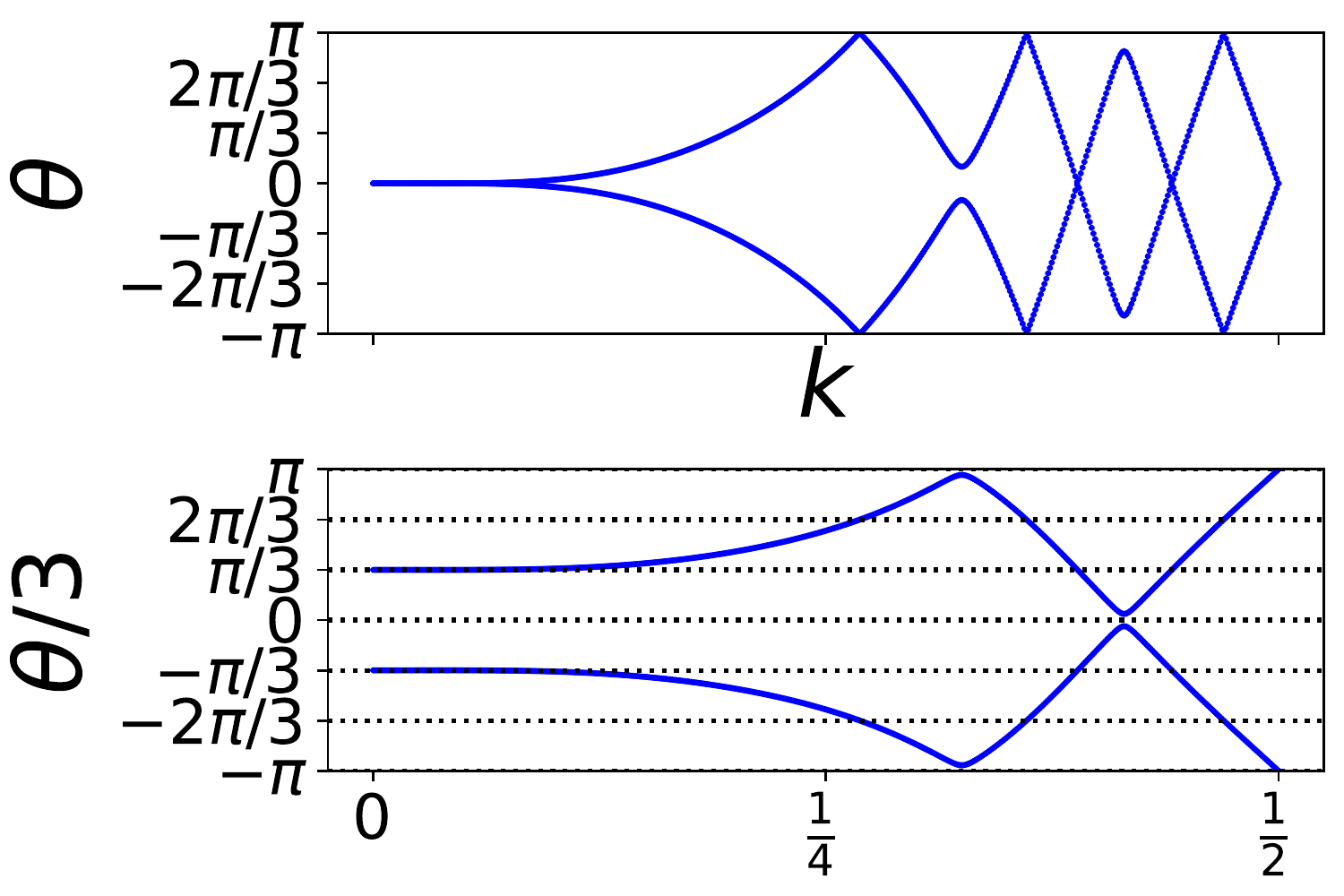}
}
\subfloat[]{
	\includegraphics[width=0.4\textwidth]{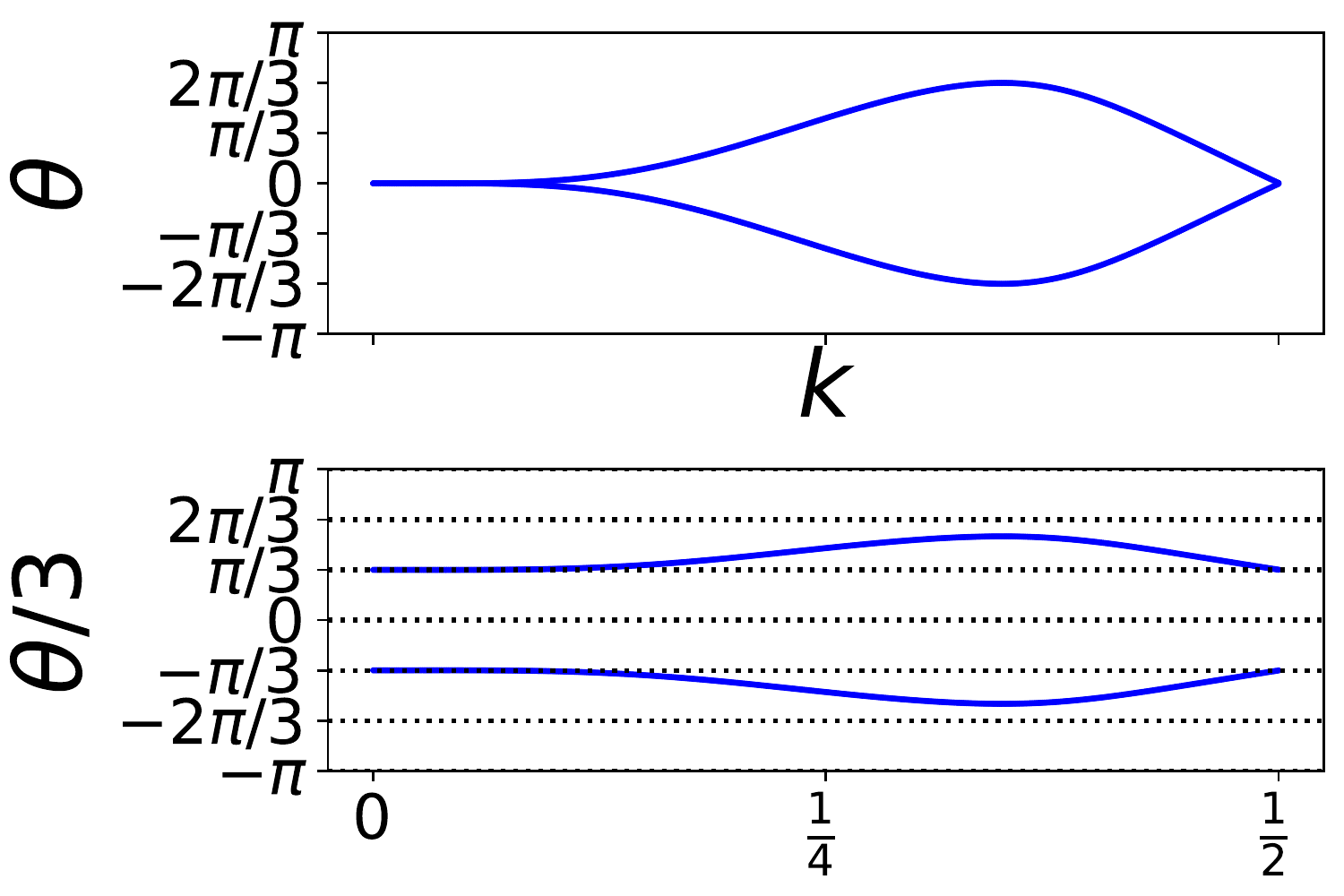}
}
\caption{Spectra of $W_h(k)$ and $W_3(k)$ for each group of two bands in the lowest four bands in our model in space group $p3m1$ without time-reversal symmetry. {Both spectra are pinned at the endpoints $k=0$ and $k=1/2$ by $C_3$ symmetry, as shown in Appendix~\ref{app:hexloopproofs}. The upper plots are obtained from those below by multiplication by three ($\mod 2\pi$), leading to crossings in the spectrum of $W_h$.} (a) shows the spectra for the middle, topological bands. The upper panel is the spectrum of $-i\log W_h(k)$, while the lower panel is the spectrum of $-i\log W_3$(k). We show in dotted lines in the lower panel the values $\theta=n\pi/3$ for integer $n$; {note the avoided crossing near $\theta/3=0,\pi$}. Upon taking the cube of this spectrum, these eigenvalues correspond to crossings in the spectrum of $W_h(k)$. The spectrum of $W_h(k)$ exhibits a nontrivial winding, reflecting the topology of these two bands. (b) Shows the same spectra for the lowest two bands. Here we see that there is no winding of the eigenvalues of $W_h(k)$.}\label{fig:hexloop3m1}
\end{figure*}

\begin{figure*}
\subfloat[]{
	\includegraphics[width=0.4\textwidth]{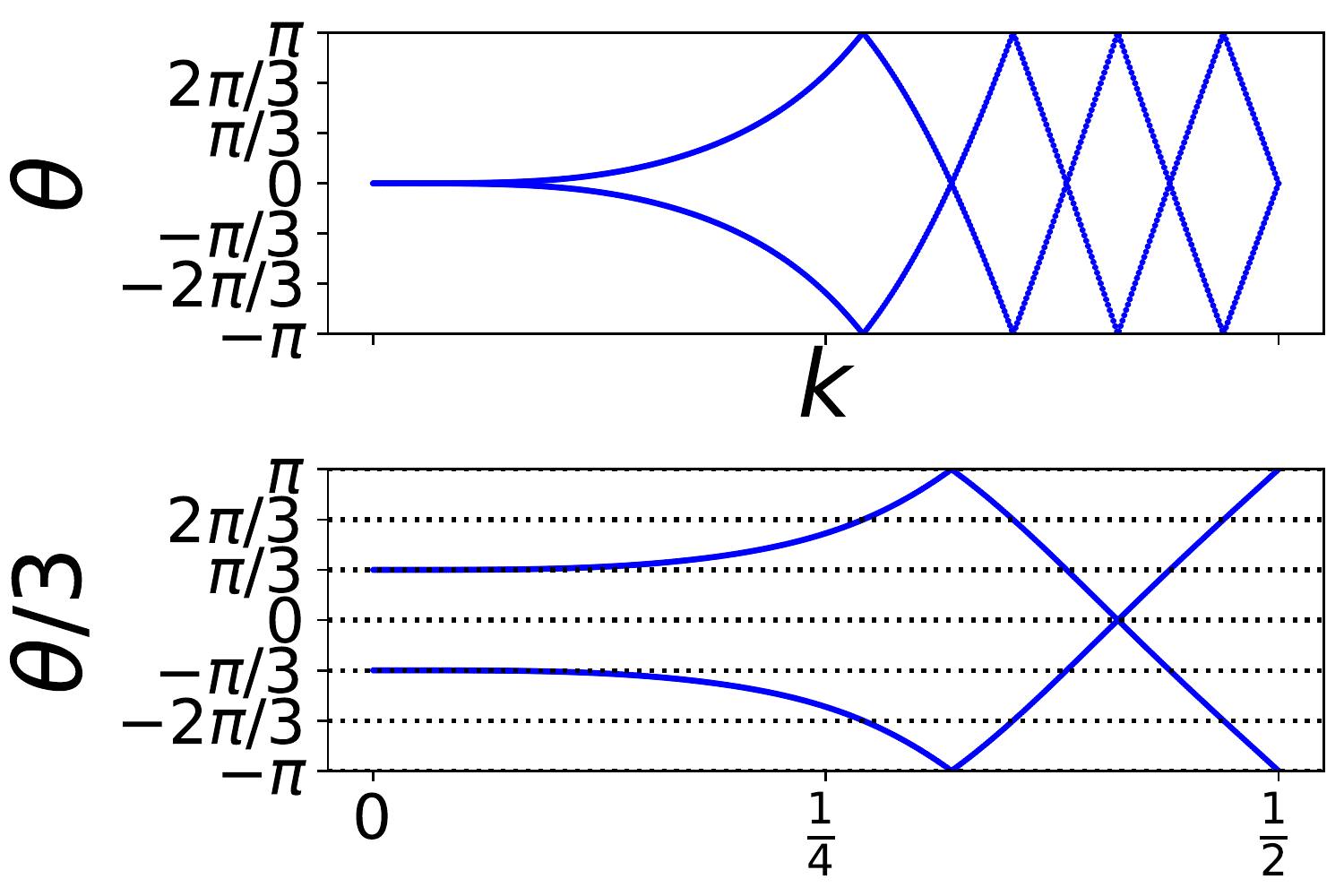}
}
\subfloat[]{
	\includegraphics[width=0.4\textwidth]{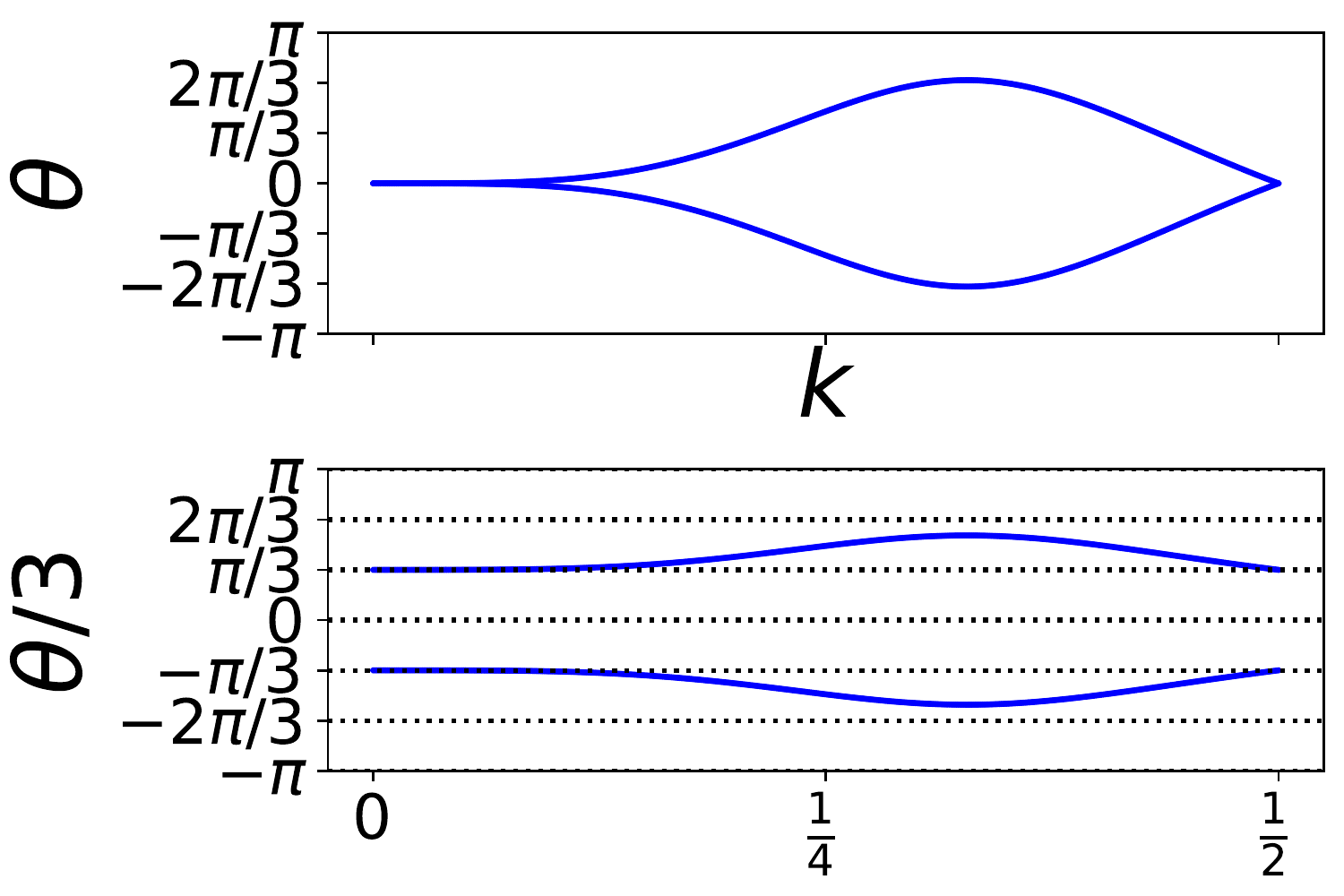}
}
\caption{Spectra of $W_h(k)$ and $W_3(k)$ for each group of two bands in the lowest four bands in our model in space group $p6mm$ with time-reversal symmetry. {Both spectra are pinned at the endpoints $k=0$ and $k=1/2$ by $C_3$ symmetry, as shown in Appendix~\ref{app:hexloopproofs}. The upper plots are obtained from those below by multiplication by three ($\mod 2\pi$), leading to crossings in the spectrum of $W_h$.} (a) shows the spectra for the middle, topological bands. The upper panel is the spectrum of $-i\log W_h(k)$, while the lower panel is the spectrum of $-i\log W_3$(k). We show in dotted lines in the lower panel the values $\theta=n\pi/3$ for integer $n$; {note that due to $C_2$ and $T$ symmetries, crossings in the spectrum of $W_3$ (and hence also $W_h$ are no longer avoided).} Upon taking the cube of this spectrum, these eigenvalues correspond to crossings in the spectrum of $W_h(k)$. The spectrum of $W_h(k)$ exhibits a nontrivial winding, reflecting the topology of these two bands. (b) Shows the same spectra for the lowest two bands. Here we see that there is no winding of the eigenvalues of $W_h(k)$.}\label{fig:hexloop6mm1prime}
\end{figure*}

As we show in Appendix~\ref{app:hexloopproofs}, {the spectrum of $W_3$ is gauge invariant owing to the fact that it is an operator which maps Bloch states at $\mathbf{k}=(k,0)$ onto themselves. Additionally, as shown in Eq.~(\ref{eq:mirrorph})} mirror symmetry forces the spectra of both $W_h(k)$ and $W_3(k)$ to be ``particle-hole'' symmetric, i.e. every eigenvalue of $e^{i\theta}$ has a partner at $e^{-i\theta}$ for $\theta \neq 0,\pi$. Furthermore, we show that for a two-band subspace, crossings in the spectrum of $W_h(k)$
are protected due to the $W_3$ (or in $p6mm$, the analogous $W_{6}$) eigenvalues of the Wilson bands. Finally, note that in the presence of time-reversal and $C_{2}$ symmetries, additional protected crossings can occur in the spectrum of a two-band hexagonal Wilson loop owing to $C_{2z}T$ symmetry (c.f.~App.~\ref{app:c2tproof}), 

Let us focus on the loop $W_h(k)$ restricted to the two-band subspace of the middle bands of our {six-band} model in both space groups $p6mm$ and $p3m1$. In Fig.~\ref{fig:hexloop6mm} we show the spectrum of $W_h$ and $W_3$ for both the {trivial (b) and topological (a) pairs of bands in space group $p6mm$, with Hamiltonian given by Eqs.~(\ref{eq:hamgen1}) and (\ref{eq:hamgen2}), with parameters given in the second column of Table~\ref{tb:paramtables}. In Fig.~\ref{fig:hexloop3m1} we show the same spectra for our model in space group $p3m1$, with parameters given in the third column of Table~\ref{tb:paramtables}. Finally, we show in Fig.~\ref{fig:hexloop6mm1prime} the same spectra in the original six-band model} \emph{with} time-reversal symmetry for comparison. We see in all cases that the spectrum of $W_h(k)$ winds nontrivially for the topological bands, showing that these bands are not adiabatically connected to an atomic limit. In these cases, the winding is guaranteed by both mirror symmeyry and the $C_3$ eigenvalues of the bands at $\Gamma$ and $K$, which forces the spectrum of $\log W_3(0)$ to be pinned at $\pm\pi/3$, while the spectrum of $\log W_3(0)$ is pinned to $\pm\pi$. Since $W_h=W_3^3$, this leads to an essential winding in the spectrum of $W_h$ (see App.~\ref{app:hexloopproofs} for the full proof).

Thus, we have here an example of a topologically nontrivial set of bands which cannot be diagnosed by a nontrivial winding along a straight Wilson loop. Nevertheless, the theory of TQC correctly predicts the nontrivial topology of this group of bands. We see then the danger of using only a small set of Wilson loops to conclude that a set of bands is topologically trivial; in general, different symmetry-distinct classes of Wilson loop, such as our hexagonal loop $W_h$ above, can reflect different obstructions to forming localized Wannier functions.

\section{Broader Outlook}\label{sec:conclusion}

The points above touch upon a more complete perspective between the theory of topological quantum chemistry beyond the conventional approach to classifying topological phases. In particular, we emphasize the importance of topological \emph{bands} and projectors, rather than focusing simply on the topology of the entire valence band manifold.
Ref.~\onlinecite{NaturePaper} is more than just a classification -- it aims not to find the allowed topological indices in a space group, but instead to enumerate all the allowed topological band connectivities; we have since tabulated this data on the Bilbao Crystallographic Server.\cite{GroupTheoryPaper,GraphDataPaper} 
Indeed, as mentioned in Ref.~\onlinecite{NaturePaper} and explored in detail in Ref.~\onlinecite{EBRTheoryPaper}, the exact elementary band representation discussed in Ref~\onlinecite{comment} also can be tuned to the
$\mathbb{Z}_2$ nontrivial phase of the Kane-Mele\cite{Kane04}, through an intermediate phase where all four bands are connected. For the same reasons as outlined above, the eigenvalue method of Ref.~\onlinecite{Po2017} cannot predict the existence of this phase.
TQC tells us such a phase can exist because the space group permits a disconnected PEBR. 
That TQC can detect \emph{both} stable (Kane-Mele model) and unstable (Eq~(\ref{eq:model})) phases within the same space group shows its power in capturing new topological physics.

Finally, we remark on the applicability of the TQC framework in light of our detailed examination of fragile phases, as well as the results of Refs.~\onlinecite{comment,Slager2018}.
Having showed that the EBR method encompasses \emph{all} the topological classes of Ref.~\onlinecite{Po2017}, and the more refined topological phases beyond K-theory, we remark that the utility of using the existence of disconnected EBRs as a tool for finding new topological materials has been established in the multiple examples presented in Refs.~\onlinecite{NaturePaper,bigmaterials,schindler2018higher,liu2018quantum}. 
The EBR method for finding materials as detailed in Refs.~\onlinecite{NaturePaper, EBRTheoryPaper, GraphTheoryPaper} is established and tried. 
Furthermore, a diagnosis of nontrivial topology on the basis of elementary band representations is more general than a case-by-case classification based on individual classes of Wilson loops, as we have shown with our discovery of the protected winding of hexagonal loops in Sec.~\ref{sec:notrhex}. Although symmetry indicators belong to an efficiently computable subset of the topological data contained in the theory of band representations, we emphasize again that the full data of a band representation is contained in the representation matrices for symmetry operations as a function of momentum, which is in-principle computable. Because Wilson loop windings like those presented here are fundamentally tied to this $\mathbf{k}$-dependence\cite{}, we suspect that an examination of more exotic disconnected EBRs will allow for the discovery of even wider varieties of topological invariants.  

\emph{Note Added:} During the completion of this work, we learned of Ref.~\onlinecite{zhidaprep}, where similar Wilson loops are considered in an unrelated context.
\begin{acknowledgements}

The authors thank A. Bouhon and R.~-J. Slager for bringing our attention to their Ref.~\onlinecite{Slager2018}, and for ensuing discussions. BB acknowledges the hospitality of the Donostia International Physics Center, where part of this work was completed. ZW and BAB were supported by the Department of Energy Grant No. DE-SC0016239, the National Science Foundation EAGER Grant No. NOA-AWD1004957, Simons Investigator Grants No. ONR-N00014-14-1-0330, No. ARO MURI W911NF-12-1-0461, and No. NSF-MRSEC DMR- 1420541, the Packard Foundation, and the Schmidt Fund for Innovative Research. ZW was supported by the CAS Pioneer Hundred Talents Program.  

\end{acknowledgements}

\appendix
\section{Review of the Models Considered}\label{app:models}
\begin{figure}
\includegraphics[width=0.4\textwidth]{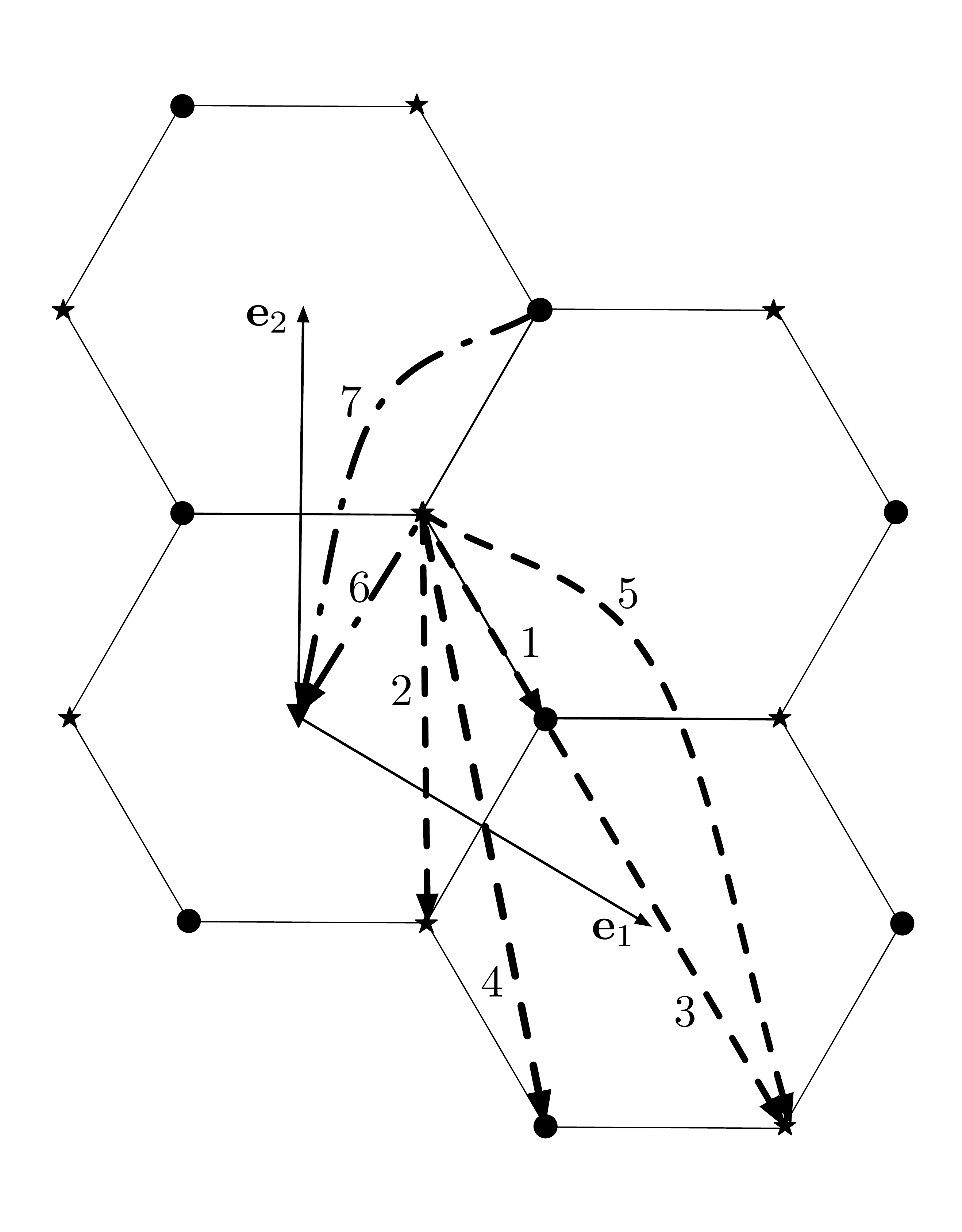}
\caption{Real-space lattice vectors and hoppings used to construct models in space groups $p3m1$, $p6mm$, and $p6mm1'$. Sites on the A sublattice are labelled with stars, those on the B sublattice with circles, and the $1a$ site at the origin of the unit cell is labelled with a triangle. Bonds used in the construction of the Hamiltonian are shown indexed from $1$--$7$. Bonds $1$--$5$ are used to construct $H_0$ and are shown as dashed lines, whereas bonds $6$ and $7$ are used to construct $H_c$ and are shown as dashed-dotted lines.}\label{fig:realspacemodel}
\end{figure}

Following Ref.~\onlinecite{Po2017}, we construct our models by choosing individual hoppings, and summing over their symmetry orbits. In particular, we consider the seven hoppings shown in Fig.~\ref{fig:realspacemodel}. For each bond $i$, ranging from $1$ to $7$ in the figure, we choose a spin-independent hopping amplitude $t_i$ and a vector of spin-orbit coupling strengths $(\lambda^x_i,\lambda_i^y,\lambda_i^z)$. We then form
\begin{equation}
h_i=\sum_\mathbf{R}(t_i\delta_{\sigma\sigma'}+i\vec{\lambda_i}\cdot\vec{\sigma}_{\sigma\sigma'})c^\dag_{if\mathbf{R}\sigma}c_{io\mathbf{R}\sigma'}\label{eq:hamgen1}
\end{equation}
where $c_{io\mathbf{R}\sigma}$ $(c_{if\mathbf{R}\sigma})$ annihilates an electron in an s (alternatively $p_z$, since under the wallpaper group symmetries considered in this paper the two are indistinguishable) of spin $\sigma$ at the origin $o$ (endpoint $f$) of the bond in unit cell $\mathbf{R}$. Note that $h_i$ is automatically time-reversal invariant provided that $t_i$ and $\vec{\lambda_i}$ are real; we break time reversal symmetry by choosing complex values for certain of these hoppings. To construct the full Hamiltonian, we employ the coset decomposition of the point group $G=6mm$,
\begin{equation}
G=\langle m_y, C_3\rangle \cup m_x\langle m_y, C_3\rangle \equiv G_0\cup m_x G_0
\end{equation}
where $m_x$ and $m_y$ are reflections that take $x\leftrightarrow -x$ and $y\leftrightarrow -y$. Here $G_0$ is isomorphic to the point group $3m$. We can thus form the Hamiltonians
\begin{align}
H_0&=\frac{1}{12}\sum_{i=1}^{5}\sum_{g\in G_0}\left(gh_ig^{-1}+x_i(m_xg)h_i(m_xg)^{-1}\right), \\
H_c&=\frac{1}{12}\sum_{i=6}^{7}\sum_{g\in G_0}\left(gh_ig^{-1}+x_i(m_xg)h_i(m_xg)^{-1}\right),\label{eq:hamgen2}
\end{align}
where $H_0$ includes all couplings between orbitals at the honeycomb lattice sites, and $H_c$ couples orbitals at the honeycomb sites to orbitals at the origin. $H_0$ and $H_c$ are combined as per Eq.~(\ref{eq:model}) to form the full Hamiltonian. We have introduced an additional set of parameters $x_i$ which control the breaking of $m_x$ symmetry, and hence the breaking of point group $6mm$ to point group $3m$; when $x_i=1$ for all $i$, the Hamiltonians $H_0$ and $H_c$ respect the full $6mm$ symmetry.

In Tables~\ref{tb:paramtables} below, we give the values of $t_i,\vec{\lambda}_i$, and $x_i$ used to construct our models in $p6mm1'$, $p6mm$, and $p3m1$ respectivey.

\begin{table*}
\begin{tabular}{c||ccccc||ccccc||ccccc}
& \multicolumn{5}{c||}{$p6mm1'$} & \multicolumn{5}{c ||}{$p6mm$} & \multicolumn{5}{c}{$p3m1$} \\
\hline
\hline
Bond & $t$ & $\lambda^x$ & $\lambda^y$ & $\lambda^z$ & $x$ & $t$ & $\lambda^x$ & $\lambda^y$ & $\lambda^z$ & $x$ & $t$ & $\lambda^x$ & $\lambda^y$ & $\lambda^z$ & $x$ \\
\hline
$1$ & $-0.7$ & $-0.4$ & $-0.2$ & $0$ & $1$ &$-0.7$ & $-0.4$ & $-0.2$ & $0$ & $1$ & $-0.7$ & $-0.4$ & $-0.2$ & $0$ & $1$ \\
$2$ & $0$ & $-0.6$ & $0$ & $-1$ & $1$ & $0$ & $-0.6$ & $0$ & $-1$ & $1$ & $0$ & $-0.6$ & $0$ & $-1$ & $0.8$ \\
$3$ & $-0.3$ & $-0.7$ & $-0.4$ & $0$ & $1$ &$-0.3$ & $-0.7$ & $-0.4$ & $0.3i$ & $1$ & $-0.3$ & $-0.7$ & $-0.4$ & $0.3i$ & $1$ \\
$4$ & $0$ & $0.9$ & $0.3$ & $0$ & $1$ &$0.1$ & $0.9$ & $0.3$ & $0.4i$ & $1$ &$0.2$ & $0.9$ & $0.3$ & $0.2i$ & $1$ \\
$5$ & $-0.2$ & $0.4$ & $0.3$ & $0.12$ & $1$ &$-0.2$ & $0.4$ & $0.3$ & $0.12$ & $1$ &$-0.2$ & $0.4$ & $0.3$ & $0.12$ & $0$\\
$6$ & $-3$ & $-2.1$ & $1.2$ & $0$ & $1$ &$-3$ & $-2.1$ & $1.2$ & $0$ & $1$ &$-3$ & $-2.1$ & $1.2$ & $0$ & $1$ \\
$7$ & $0.5$ & $1.2$ & $-0.7$ & $0$ & $1$ &$0.5$ & $1.2$ & $-(0.7+1.4i)$ & $0$ & $1$ & $0.5$ & $1.2$ & $-(0.7+1.4i)$ & $0$ & $1$ \\
\end{tabular}
\caption{Values of the parameters $t_i,\vec{\lambda}_i$ and $x_i$ used to construct the models discussed in the main text. The definition of these parameters is given in Eqs.~(\ref{eq:hamgen1}--\ref{eq:hamgen2}). Note the complex parameter values used to break time-reversal symmetry, and the values of $x\neq 1$ used to break $C_6$ symmetry.}\label{tb:paramtables}
\end{table*}

\section{Generic Wilson loop crossings}\label{app:c2tproof}

In this Appendix we will show that $C_{2z}$ and $T$ symmetry allow for and stabilize crossings in the two-band Wilson loop, away from high-symmetry $\mathbf{k}$-points. Let us consider the family of $\mathbf{k}_\parallel$-oriented Wilson loop matrices $W(\mathbf{k}_\perp)$, parametrized by $\mathbf{k}_\perp$. While a convenient choice is to orient $\mathbf{k}_\parallel$ and $\mathbf{k}_\perp$ along the basis vectors $\mathbf{g}_1$ and $\mathbf{g}_2$ of the reciprocal lattice (as in Fig.~\ref{fig:0muwilsoncond}, this is not necessary for the proof that follows). Now, both $C_{2z}$ and time-reversal symmetry take $\mathbf{k}_\parallel\rightarrow-\mathbf{k}_\parallel$ as well as $\mathbf{k}_\perp\rightarrow-\mathbf{k}_\perp$. This implies that
\begin{align}
\mathcal{T}W(\mathbf{k}_\perp)\mathcal{T}^{-1}&=W^\dag(-\mathbf{k}_\perp)\\
C_{2z}W(\mathbf{k}_\perp)C_{2z}^{-1}&=W^\dag(-\mathbf{k}_\perp),
\end{align}
where $\mathcal{T}$ is the antiunitary operator that implements $T$-symmetry. Defining the Wilson Hamiltonian $\mathcal{H}_W(\mathbf{k}_\perp)$ by
\begin{equation}
W(\mathbf{k}_\perp)\equiv e^{i\mathcal{H}_W(\mathbf{k}_\perp)},\label{eq:wilsonham}
\end{equation}
we deduce
\begin{align}
\mathcal{T}\mathcal{H}_W(\mathbf{k}_\perp)\mathcal{T}^{-1}&=\mathcal{H}_W(-\mathbf{k}_\perp) \label{eq:tconstraint}\\
C_{2z}\mathcal{H}_W(\mathbf{k}_\perp)C_{2z}^{-1}&=-\mathcal{H}_W(-\mathbf{k}_\perp),\label{eq:c2constraint}
\end{align}
where the antiunitarity of $\mathcal{T}$ accounts for the additional minus sign. We also have the combined symmetry
\begin{equation}
C_{2z}\mathcal{T}\mathcal{H}_W(\mathbf{k}_\perp)(C_{2z}\mathcal{T})^{-1}=-\mathcal{H}_W(\mathbf{k}_\perp) \mod 2\pi.\label{eq:phconstraint}
\end{equation}
We note briefly that this equation {permits solutions with $\mathbf{k}_\perp$-independent gapped spectrum $\mathrm{spec}[\mathcal{H}_W(\mathbf{k}_\perp)]=\{0,\pi\}$}. However, since we are mostly concerned with time-reversal symmetric gapped systems (where this solution is ruled out by Kramers's theorem), we will neglect this possibility.

We will now examine the algebraic constraints placed on $2\times2$ matrix representatives of these symmetries in the two-band subspace [the image of the projectors $P$ in Eq.~(\ref{eq:wilsondef})], and from these deduce restrictions on the form of $\mathcal{H}_W$. Unitarity implies that we can chose representatives
\begin{align}
\mathcal{T}&=U_{T}\mathcal{K} \\
C_{2z}&=U_{2},
\end{align}
where $U_T$ and $U_2$ are unitary, and $\mathcal{K}$ is the complex conjugation operation. The constraints $\mathcal{T}^2=C_{2z}^2=-\mathbb{I}$, along with $\mathcal{T}C_{2z}\mathcal{T}^{-1}=C_{2z}$ translate to
\begin{align}
U_2^2&=-\sigma_0,\\
U_TU_T^*&=-\sigma_0 \\
U_2U_T&=U_TU_2^*,
\end{align}
where we have introduced and will use $(\sigma_0,\vec{\sigma})$ for the identity matrix and the vector of three Pauli matrices $\sigma_x,\sigma_y,$ and $\sigma_z$ respectively. The most general solution to these constraints is given by
\begin{align}
U_T&=iU(\mathbf{k}_\perp)\sigma_yU(\mathbf{k}_\perp)^T,\\
U_2&=iU(\mathbf{k}_\perp)(\mathbf{\hat{n}}\cdot\vec{\sigma})U(\mathbf{k}_\perp)^\dag,
\end{align}
where $\mathbf{\hat{n}}$ is an arbitrary unit vector, and $U(\mathbf{k}_\perp)$ is an arbitrary (smooth) unitary matrix. Inserting these into the constraint Eq.~(\ref{eq:phconstraint}), we see that
\begin{equation}
\mathcal{H}_W(\mathbf{k}_\perp)=a(\mathbf{k}_\perp)U(\mathbf{k}_\perp)(\mathbf{\hat{n}}\cdot\vec{\sigma})U(\mathbf{k}_\perp)^\dag.
\end{equation}
Inserting this expression into the definition Eq.~(\ref{eq:wilsonham}), we see that there is a crossing in the Wilson loop spectrum whenever $a(\mathbf{k}_\perp)=n\pi$ for $n\in\mathbb{Z}$; for $n$ even these crossings occur at $\theta=i\log W=0$, and for $n$ odd they occur at $\theta=\pm\pi$. Furthermore, since the condition for a crossing depends on the value of a single function of one variable (in two dimensions), linear crossings in the Wilson loop spectrum can only be created or destroyed in pairs. We thus conclude that the Wilson loop windings in Figs.~\ref{fig:0muwilsoncond} and \ref{fig:topwilson08} are robust.

\section{Hexagonal Wilson loops}\label{app:hexloopproofs}

While the fragile phases in space group $p6mm1'$ mentioned above can be accurately diagnosed by the $C_2T$ protected winding of straight Wilson loops, neither $C_2$ nor time-reversal symmetry are necessary to protect this topological phase. In fact, our EBR analysis above showed that the nontrivial nature of this phase can be diagnosed solely by comparing the $C_3$ eigenvalues at the $K,K'$ and $\Gamma$ points to the table of EBRs. {Furthermore, this same set of eigenvalues is also incompatible with any two-band band representation
in the space groups $p6mm$, $p3m11'$, and $p3m1$ (obtained from $p6mm1'$ by breaking time reversal, $C_2$, and both $C_2$ and time reversal, respectively). Hence $C_3$ and mirror symmetry alone are enough to guarantee that these two bands are topologically nontrivial.}
This suggests we look for a diagnostic of this topological phase dependent on $C_3$ and mirror symmetries.

To do so, let us examine the concrete case of $p3m1$. We take the same direct and reciprocal lattice vectors as in the main text,
\begin{align}
\mathbf{e}_1&=\frac{\sqrt{3}}{2}\mathbf{\hat{x}}-\frac{1}{2}\mathbf{\hat{y}} \\
\mathbf{e}_2&=\mathbf{\hat{y}}
\end{align}
and
\begin{align}
\mathbf{g}_1&=\frac{4\pi}{\sqrt{3}}\mathbf{\hat{x}} \\
\mathbf{g}_2&=2\pi\left(\frac{1}{\sqrt{3}}\mathbf{\hat{x}}+\mathbf{\hat{y}}\right)
\end{align}

Recall that with this choice of lattice vectors, the high symmetry points have the reduced coordinates
\begin{align}
K&=\left(\frac{1}{3},\frac{1}{3}\right), \\
K'&=\left(-\frac{1}{3},\frac{2}{3}\right), \\
M&=\left(\frac{1}{2},0\right), \\
M'&=\left(-\frac{1}{2},\frac{1}{2}\right), \\
M''&=\left(0,\frac{1}{2}\right),
\end{align}
{modulo reciprocal lattice translations}. For what follows the relevant symmetries are $C_3$, and $m_y$ {(we will alson briefly comment on the role of time-reversal symmetry)}. We recall that $m_y$ interchanges $K$ and $K'$, and leaves $M$ invariant.

To define our topological invariant, consider the wilson line $W_\mathcal{C}(k)$ evaluated along the following path:
\begin{equation}
\mathcal{C}=\{(k,0)\rightarrow (2k/3,2k/3)\rightarrow (-2k/3,4k/3)\rightarrow(-k,k)\}.
\end{equation}
We emphasize that $W_\mathcal{C}(k)$ has $(k,0)$ as a basepoint. The path $\mathcal{C}$ is one third of a hexagon encircling the $\Gamma$ point. We can thus define the full hexagonal Wilson loop as
\begin{align}
W_h(k)&=W_{C_3^2\mathcal{C}}(k)W_{C_3\mathcal{C}}(k)W_\mathcal{C}(k) \\
&= C_3^2W_\mathcal{C}(k)C_3^{-2}C_3W_\mathcal{C}(k)C_3^{-1}W_\mathcal{C}(k) \\
&= -(C_3^{-1}W_\mathcal{C}(k))^3
\end{align}
where we have used the fact that $C_3^2=-C_3^{-1}$. Note that the above equation is a statement about \emph{operators} on the space of Bloch functions. In particular, we recognize 
\begin{equation}
W_3 (k)\equiv C_3^{-1}W_\mathcal{C}(k)
\end{equation}
as (the inverse of) the operator which implements $C_3$ symmetry on the loop $W_h$(see, for instance, Ref.~\onlinecite{ArisCohomology}), i.e.
\begin{align}
C_3W_h(k)C_3^{-1}&=C_3(C_3^{-1}W_\mathcal{C}(k))^3C_3^{-1} \\
&=W_\mathcal{C}(k)(C_3^{-1}W_\mathcal{C}(k))^3W_\mathcal{C}^{-1}
\\&=W_\mathcal{C}(k)W_h(k)W_\mathcal{C}(k)^{-1},
\end{align}
where the effect of the Wilson line $W_{\mathcal{C}}(k)$ is to account for the fact that a $C_3$ rotation shifts the basepoint of $W_h(k)$ from $(k,0)$ to $(-k,k)$. We have thus concocted an interesting situation where a Wilson loop $W_h(k)$ is expressible as a function of the generator of a symmetry of the loop, i.e.,
\begin{equation}
[W_h(k),W_3 (k)]=0,\;\; W_h(k)=-W_3 (k)^3.\label{eq:cubereln}
\end{equation}

Let us next examine the matrix elements of these operators in a subspace of bands $\{|u_\mathbf{k}^m>\}$. First, we note that $W_\mathcal{C}(k)$ has nonvanishing matrix elements between states $|u^m_{(k,0)}\rangle$ and $\langle u^n_{(-k,k)}|$, while $C_3$ has nonvanishing matrix elements between $|u^m_\mathbf{k}\rangle$ and $\langle u^n_{C_3\mathbf{k}}|$. {Neither of these operators individually has a gauge-invariant spectrum. However,} putting them together, we have that
\begin{equation}
\left(W_3 \right)^{mn}(k)=\langle u^m_{(k,0)}|C_3^{-1}|u^\ell_{(-k,k)}\rangle\langle u^\ell_{(-k,k)}|W(k)|u^n_{(k,0)}\rangle
\end{equation}
transforms covariantly under unitary transformations in this subspace of bands -- under a $\mathbf{k}$-dependent unitary transformation, $\left(W_3 \right)^{mn}(k)$ is conjugated by the unitary evaluated at $(k,0)$. Thus, $W_3$ has gauge invariant eigenvalues. {Furthermore, since $[W_3(k),W_h(k)]=0$, the two can be simultaneously diagonalized.
We can thus label the Wilson bands of $W_h(k)$ by their eigenvalues under $W_3(k)$; the interpretation is that the dressed Wilson line $W_3(k)$ implements a $C_3$ rotation on the eigenstates of $W_h(k)$.}

Let us now examine some specific properties of $W_3(k)$ and $W_h(k)$. First, let us rewrite $W_3(k)$ as a product of Wilson lines between sides and corners of the hexagon,

\begin{align}
W_3(k)=&C_3^{-1}W_{(-k,k)\leftarrow (-2k/3,4k/3)}\times\nonumber \\
&\times W_{(-2k/3,4k/3)\leftarrow(2k/3,2k/3)}W_{(2k/3,2k/3)\leftarrow(k,0)},
\end{align}
from which it follows that
\begin{widetext}
\begin{align}
m_yW_3(k)m_y^{-1}&=C_3W_{(0,-k)\leftarrow (2k/3,-4k/3)}W_{(2k/3,-4k/3)\leftarrow(4k/3,-2k/3)}W_{(4k/3,-2k/3)\leftarrow(k,0)}, \nonumber \\
&=W_{(k,0)\leftarrow (2k/3,2k/3)}W_{(2k/3,2k/3)\leftarrow(-2k/3,4k/3)}W_{(-2k/3,4k/3)\leftarrow(-k,k)}C_3,\nonumber \\
&=W_3(k)^{-1},\label{eq:mirrorph}
\end{align}
\end{widetext}
where we have used the group relation $m_yC_3m_y^{-1}=C_3^{-1}$, along with the action of $C_3$ on bloch functions and wavevectors. From this, it follows that the spectrum of $W_3(k)$ is ``particle-hole'' symmetric, and consequently the same for the spectrum of $W_h(k)$.

We will use the holonomies $W_3(k)$ and $W_h(k)$ to show the nontriviality (i.e. nonWannierizability) of various topological bands. The crux of the argument will be to compare $W_h(\Gamma)$ (i.e.~$W_h(k=0)$) to $W_h(M)$ (i.e.~$W_h(1/2)$), and deduce that there must be an unremovable winding in $W_h(k)$. Since the path along which $W_h(k)$ is computed covers the whole Brillouin zone as $k$ is varied, we will deduce that a group of bands is not homotopic to the atomic limit. To begin, we note first that for $k=0$ we have
\begin{align}
W_3(0)&=C_3^{-1}(\Gamma) \\
W_h(0)&=P(\Gamma),
\end{align}
where $P(\mathbf{k})$ is defined as the projector onto the bands of interest. Evaluating these matrices in the subspace of the image of $P(\mathbf{k})$ gives
\begin{align}
\left(W_3\right)^{mn}(0)&=B_{C_3^{-1}}^{mn}(\Gamma) \label{eq:c30} \\
W_h^{mn}(0)&=\delta_{mn}
\end{align}
where $B_g(\mathbf{k})$ denotes the sewing matrix\cite{Fang2012,ArisCohomology} for the symmetry $g$ at the point $\mathbf{k}$.

Next, we examine $W_3\left(1/2\right)$, {which is the line $MKM''K'M'$. It is helpful to first rewrite
\begin{align}
W_3\left(\frac{1}{2}\right)&=C_3^{-1}W_{M'\leftarrow K'}W_{K'\leftarrow M''\leftarrow K}W_{K\leftarrow M} \\
&=W_{M\leftarrow(K'-\mathbf{g}_2+\mathbf{g}_1)}C_3C_3^{-2}W_{K'\leftarrow M''\leftarrow K}W_{K\leftarrow M} \\
&=W_{M\leftarrow(K'-\mathbf{g}_2+\mathbf{g}_1)}C_3W_{(K'-\mathbf{g}_2)\leftarrow (M-\mathbf{g}_1)\leftarrow K-\mathbf{g}_1}\times\nonumber \\
&\times C_3^{-2}W_{K\leftarrow M}
\end{align}
Conjugating by $W_{K\leftarrow M}$ and using $C_3^{-2}=-C_3$, we find
\begin{widetext}
\begin{align}
W_{K\leftarrow M}W_3(1/2)W_{M\leftarrow K}&=-W_{K\leftarrow M\leftarrow (K'-\mathbf{g}_2+\mathbf{g}_1)}C_3W_{(K'-\mathbf{g}_2)\leftarrow (M-\mathbf{g}_1)\leftarrow K-\mathbf{g}_1}C_3\\
&=-W_{K\leftarrow M\leftarrow (K'-\mathbf{g}_2+\mathbf{g}_1)}C_3V(\mathbf{g}_1)W_{(K'-\mathbf{g}_2+\mathbf{g}_1)\leftarrow (M)\leftarrow K}V^\dag(\mathbf{g}_1)C_3,
\end{align}
where $V(\mathbf{g}_1)$ is the unitary matrix relating Bloch states in adjacent Brillouin zones. Introducing the shorthand $K'_{12}=K'+\mathbf{g}_1-\mathbf{g}_2$ and taking matrix elements remembering the periodicity of the sewing matrices, we have thus
\begin{equation}
(W_{KM}W_3(1/2)W^\dag_{KM})^{mn}=-W_{KMK'_{12}}^{m\ell}B_{C_3}^{\ell r}(K')W_{K'_{12}MK}^{rs}B_{C_3}^{sn}(K)\label{eq:c3half}
\end{equation}
\end{widetext}
}
This shows that $W_3$, as a matrix in the space of bands under consideration, is unitarily equivalent to (minus) a product of the $C_3$ sewing matrix at $K$, and the $C_3$ sewing matrix at $K'$ parallel transported to $K$.

Lastly, let us examine the role of $W_3$ eigenvalues $e^{i\theta/3}$ on the spectrum of $W_h(k)$. Using Eq.~\ref{eq:cubereln}, we see that each pair of $\theta/3=\pm n\pi/3$ eigenvalues of $W_3$ leads to a pair of Wilson bands crossing with $W_h$ eigenvalues $\theta=(n+1)\pi$. Furthermore, provided $n\neq 0 \mod 3$, these crossings are protected, as they correspond to an intersection of Wilson bands with different $W_3$ eigenvalues. Absent additional symmetries (see, for example, below), crossings with $n=0\mod 3$ are not stable. Note that while these statements use the mirror-enforced particle-hole symmetry of the Wilson lines $W_3$ and $W_h$, they hold irrespective of the number of bands in which the Wilson loop is evaluated (the rank of the projector).

Note that in systems with additional $C_6$ symmetry, we can define an analogous operation
\begin{equation}
W_{6}=C_6^{-1}W_{(0,k)\leftarrow(k,k)\leftarrow(k,0)}
\end{equation}
which satisfies
\begin{equation}
W_3=(W_{6})^2,
\end{equation}
Which allows us to further label the eigenstates of $W_3$ and $W_h$ by their $W_{6}$ eigenvalues. As above, $m_y$ symmetry imposes particle-hole symmetry on the spectrum of $W_{6}$. In particular, states with $W_{6}$ eigenvalues given by $\pm i$ have $W_3$ eigenvalues with $\theta/3=\pm \pi$. Thus, the addition of $C_6$ symmetry stabilizes crossings in the $W_3$ spectrum with $n=6m+3$, as seen for instance in Fig.~\ref{fig:hexloop6mm}. {This leads to additional protected crossings in the spectrum of $W_h$ at $\theta=\pi$.}

With these pieces in place, let us examine the fragile topological phase introduced in the main text. We now specialize to the case where the Wilson loops are defined with rank-2 projectors, i.e.~the case of two isolated bands. The topologically nontrivial bands from Sec.~\ref{sec:model} have for their $C_3$ sewing matrices\cite{GroupTheoryPaper,Po2017}
\begin{align}
B_{C_3}(\Gamma)&=e^{-i\pi\sigma_z/3}\nonumber \\
B_{C_3}(K)&=B_{C_3}(K')=-\sigma_0\label{eq:c3sewing}
\end{align}
Inserting these into Eqs.~(\ref{eq:c30}) and (\ref{eq:c3half}) above, {and using the fact that 
\begin{equation}
W^{mn}_{KMK'_{12}}W^{n\ell}_{K'_{12}MK}=\sigma_0^{m\ell},
\end{equation}
}
 we find that
\begin{equation}
W_3(0)=e^{i\pi\sigma_z/3}, \;\; W_3\left(\half\right)=-\sigma_0,
\end{equation}
i.e. the eigenvalues of $W_3(0)$ are $e^{\pm i\pi/3}$, and the eigenvalues of $W_3\left(\half\right)$ are $(-1,-1)$. First, this implies that
\begin{equation}
W_h(0)=W_h\left(\half\right)=\sigma_0,
\end{equation}
so that at $k=0$ and $k=1/2$ the spectrum of $\log W_h$ is pinned to zero. Furthermore, continuity and particle-hole symmetry of the spectrum of $W_3$ implies that the eigenvalues of $W_3$ must pass through $e^{\pm 2i\pi/3}$ and odd number of times. Each time a pair of bands passes through $e^{\pm 2i\pi/3}$, we get a crossing in the spectrum of $\log W_h$ at $\pm \pi$. We thus deduce that as we tune $k$ from $0$ to $\half$, the spectrum of $\log W_h$ starts at $(0,0)$, goes through $\pm \pi$ and odd number of times, and returns to $(0,0)$. Thus, there is necessarily a winding of the spectrum of $W_h$, and we deduce that this group of bands cannot be described by exponentially localized, symmetric Wannier functions.

We note that the essential ingredients to prove the necessity of this Wilson loop winding were $C_3$ symmetry, and particle-hole symmetry of the Wilson loop spectrum. While in our concrete example we used mirror to enforce particle-hole symmetry, time reversal also has the same effect. Thus, in any space group which contains either $p3m1$ or $p31'$ as a subgroup, a group of bands with the combination of $C_3$ sewing matrices given in Eq.~(\ref{eq:c3sewing}) must be topologically nontrivial. Furthermore, even in space group $p3$ {(with or without time-reversal symmetry)}, such bands \emph{can} be topological, although the winding is not mandated. Nevertheless, ``adiabatic'' breaking of mirror or time-reversal symmetry will lead to a topolpogical phase, a la inversion symmetry breaking in the Kane-Mele model.

We see then that ``symmetry-indicated'' fragile topological bands, i.e.~topological bands whose little group irreps are equivalent to a difference of EBRs, can manifest obstructions to Wannierizability in unconventional Wilson loops. In general then, the criteria of whether or not a group of bands originated from a disconnected elementary band representation provides the most direct indication of an obstruction to Wannierizability. We have thus also shown that the \emph{unstable} equivariant homotopy theory of Bloch bundles holds many surprises for future work.
\bibliography{connectivity}

\begin{thebibliography}{35}%
\makeatletter
\providecommand \@ifxundefined [1]{%
 \@ifx{#1\undefined}
}%
\providecommand \@ifnum [1]{%
 \ifnum #1\expandafter \@firstoftwo
 \else \expandafter \@secondoftwo
 \fi
}%
\providecommand \@ifx [1]{%
 \ifx #1\expandafter \@firstoftwo
 \else \expandafter \@secondoftwo
 \fi
}%
\providecommand \natexlab [1]{#1}%
\providecommand \enquote  [1]{``#1''}%
\providecommand \bibnamefont  [1]{#1}%
\providecommand \bibfnamefont [1]{#1}%
\providecommand \citenamefont [1]{#1}%
\providecommand \href@noop [0]{\@secondoftwo}%
\providecommand \href [0]{\begingroup \@sanitize@url \@href}%
\providecommand \@href[1]{\@@startlink{#1}\@@href}%
\providecommand \@@href[1]{\endgroup#1\@@endlink}%
\providecommand \@sanitize@url [0]{\catcode `\\12\catcode `\$12\catcode
  `\&12\catcode `\#12\catcode `\^12\catcode `\_12\catcode `\%12\relax}%
\providecommand \@@startlink[1]{}%
\providecommand \@@endlink[0]{}%
\providecommand \url  [0]{\begingroup\@sanitize@url \@url }%
\providecommand \@url [1]{\endgroup\@href {#1}{\urlprefix }}%
\providecommand \urlprefix  [0]{URL }%
\providecommand \Eprint [0]{\href }%
\providecommand \doibase [0]{http://dx.doi.org/}%
\providecommand \selectlanguage [0]{\@gobble}%
\providecommand \bibinfo  [0]{\@secondoftwo}%
\providecommand \bibfield  [0]{\@secondoftwo}%
\providecommand \translation [1]{[#1]}%
\providecommand \BibitemOpen [0]{}%
\providecommand \bibitemStop [0]{}%
\providecommand \bibitemNoStop [0]{.\EOS\space}%
\providecommand \EOS [0]{\spacefactor3000\relax}%
\providecommand \BibitemShut  [1]{\csname bibitem#1\endcsname}%
\let\auto@bib@innerbib\@empty
\bibitem [{\citenamefont {Bradlyn}\ \emph {et~al.}(2017)\citenamefont
  {Bradlyn}, \citenamefont {Elcoro}, \citenamefont {Cano}, \citenamefont
  {Vergniory}, \citenamefont {Wang}, \citenamefont {Felser}, \citenamefont
  {Aroyo},\ and\ \citenamefont {Bernevig}}]{NaturePaper}%
  \BibitemOpen
  \bibfield  {author} {\bibinfo {author} {\bibfnamefont {B.}~\bibnamefont
  {Bradlyn}}, \bibinfo {author} {\bibfnamefont {L.}~\bibnamefont {Elcoro}},
  \bibinfo {author} {\bibfnamefont {J.}~\bibnamefont {Cano}}, \bibinfo {author}
  {\bibfnamefont {M.~G.}\ \bibnamefont {Vergniory}}, \bibinfo {author}
  {\bibfnamefont {Z.}~\bibnamefont {Wang}}, \bibinfo {author} {\bibfnamefont
  {C.}~\bibnamefont {Felser}}, \bibinfo {author} {\bibfnamefont {M.~I.}\
  \bibnamefont {Aroyo}}, \ and\ \bibinfo {author} {\bibfnamefont {B.~A.}\
  \bibnamefont {Bernevig}},\ }\href@noop {} {\bibfield  {journal} {\bibinfo
  {journal} {Nature}\ }\textbf {\bibinfo {volume} {547}},\ \bibinfo {pages}
  {298} (\bibinfo {year} {2017})}\BibitemShut {NoStop}%
\bibitem [{\citenamefont {Zak}(1980)}]{Zak1980}%
  \BibitemOpen
  \bibfield  {author} {\bibinfo {author} {\bibfnamefont {J.}~\bibnamefont
  {Zak}},\ }\href {\doibase 10.1103/PhysRevLett.45.1025} {\bibfield  {journal}
  {\bibinfo  {journal} {Phys. Rev. Lett.}\ }\textbf {\bibinfo {volume} {45}},\
  \bibinfo {pages} {1025} (\bibinfo {year} {1980})}\BibitemShut {NoStop}%
\bibitem [{\citenamefont {Zak}(1981)}]{Zak1981}%
  \BibitemOpen
  \bibfield  {author} {\bibinfo {author} {\bibfnamefont {J.}~\bibnamefont
  {Zak}},\ }\href@noop {} {\bibfield  {journal} {\bibinfo  {journal} {Phys.
  Rev. B}\ }\textbf {\bibinfo {volume} {23}},\ \bibinfo {pages} {2824}
  (\bibinfo {year} {1981})}\BibitemShut {NoStop}%
\bibitem [{\citenamefont {H\"oller}\ and\ \citenamefont
  {Alexandradinata}(2018)}]{aris18-1}%
  \BibitemOpen
  \bibfield  {author} {\bibinfo {author} {\bibfnamefont {J.}~\bibnamefont
  {H\"oller}}\ and\ \bibinfo {author} {\bibfnamefont {A.}~\bibnamefont
  {Alexandradinata}},\ }\href {\doibase 10.1103/PhysRevB.98.024310} {\bibfield
  {journal} {\bibinfo  {journal} {Phys. Rev. B}\ }\textbf {\bibinfo {volume}
  {98}},\ \bibinfo {pages} {024310} (\bibinfo {year} {2018})}\BibitemShut
  {NoStop}%
\bibitem [{\citenamefont {Po}\ \emph {et~al.}(2017)\citenamefont {Po},
  \citenamefont {Vishwanath},\ and\ \citenamefont {Watanabe}}]{Po2017}%
  \BibitemOpen
  \bibfield  {author} {\bibinfo {author} {\bibfnamefont {H.~C.}\ \bibnamefont
  {Po}}, \bibinfo {author} {\bibfnamefont {A.}~\bibnamefont {Vishwanath}}, \
  and\ \bibinfo {author} {\bibfnamefont {H.}~\bibnamefont {Watanabe}},\
  }\href@noop {} {\bibfield  {journal} {\bibinfo  {journal} {Nat. Comm.}\
  }\textbf {\bibinfo {volume} {8}},\ \bibinfo {pages} {50} (\bibinfo {year}
  {2017})}\BibitemShut {NoStop}%
\bibitem [{\citenamefont {Shiozaki}\ \emph {et~al.}(2017)\citenamefont
  {Shiozaki}, \citenamefont {Sato},\ and\ \citenamefont {Gomi}}]{Shiozaki2017}%
  \BibitemOpen
  \bibfield  {author} {\bibinfo {author} {\bibfnamefont {K.}~\bibnamefont
  {Shiozaki}}, \bibinfo {author} {\bibfnamefont {M.}~\bibnamefont {Sato}}, \
  and\ \bibinfo {author} {\bibfnamefont {K.}~\bibnamefont {Gomi}},\ }\href@noop
  {} {\bibfield  {journal} {\bibinfo  {journal} {Phys. Rev. B}\ }\textbf
  {\bibinfo {volume} {95}},\ \bibinfo {pages} {235425} (\bibinfo {year}
  {2017})}\BibitemShut {NoStop}%
\bibitem [{\citenamefont {Freed}\ and\ \citenamefont
  {Moore}(2013)}]{Freed2013}%
  \BibitemOpen
  \bibfield  {author} {\bibinfo {author} {\bibfnamefont {D.~S.}\ \bibnamefont
  {Freed}}\ and\ \bibinfo {author} {\bibfnamefont {G.~W.}\ \bibnamefont
  {Moore}},\ }\href {\doibase 10.1007/s00023-013-0236-x} {\bibfield  {journal}
  {\bibinfo  {journal} {Annales Henri Poincar{\'e}}\ }\textbf {\bibinfo
  {volume} {14}},\ \bibinfo {pages} {1927} (\bibinfo {year}
  {2013})}\BibitemShut {NoStop}%
\bibitem [{\citenamefont {Kruthoff}\ \emph {et~al.}(2017)\citenamefont
  {Kruthoff}, \citenamefont {de~Boer}, \citenamefont {van Wezel}, \citenamefont
  {Kane},\ and\ \citenamefont {Slager}}]{Kruthoff2016}%
  \BibitemOpen
  \bibfield  {author} {\bibinfo {author} {\bibfnamefont {J.}~\bibnamefont
  {Kruthoff}}, \bibinfo {author} {\bibfnamefont {J.}~\bibnamefont {de~Boer}},
  \bibinfo {author} {\bibfnamefont {J.}~\bibnamefont {van Wezel}}, \bibinfo
  {author} {\bibfnamefont {C.~L.}\ \bibnamefont {Kane}}, \ and\ \bibinfo
  {author} {\bibfnamefont {R.-J.}\ \bibnamefont {Slager}},\ }\href@noop {}
  {\bibfield  {journal} {\bibinfo  {journal} {Phys. Rev. X}\ }\textbf {\bibinfo
  {volume} {7}},\ \bibinfo {pages} {041069} (\bibinfo {year}
  {2017})}\BibitemShut {NoStop}%
\bibitem [{\citenamefont {Elcoro}\ \emph {et~al.}(2017)\citenamefont {Elcoro},
  \citenamefont {Bradlyn}, \citenamefont {Wang}, \citenamefont {Vergniory},
  \citenamefont {Cano}, \citenamefont {Felser}, \citenamefont {Bernevig},
  \citenamefont {Orobengoa}, \citenamefont {de~la Flor},\ and\ \citenamefont
  {Aroyo}}]{GroupTheoryPaper}%
  \BibitemOpen
  \bibfield  {author} {\bibinfo {author} {\bibfnamefont {L.}~\bibnamefont
  {Elcoro}}, \bibinfo {author} {\bibfnamefont {B.}~\bibnamefont {Bradlyn}},
  \bibinfo {author} {\bibfnamefont {Z.}~\bibnamefont {Wang}}, \bibinfo {author}
  {\bibfnamefont {M.~G.}\ \bibnamefont {Vergniory}}, \bibinfo {author}
  {\bibfnamefont {J.}~\bibnamefont {Cano}}, \bibinfo {author} {\bibfnamefont
  {C.}~\bibnamefont {Felser}}, \bibinfo {author} {\bibfnamefont {B.~A.}\
  \bibnamefont {Bernevig}}, \bibinfo {author} {\bibfnamefont {D.}~\bibnamefont
  {Orobengoa}}, \bibinfo {author} {\bibfnamefont {G.}~\bibnamefont {de~la
  Flor}}, \ and\ \bibinfo {author} {\bibfnamefont {M.~I.}\ \bibnamefont
  {Aroyo}},\ }\href@noop {} {\bibfield  {journal} {\bibinfo  {journal} {J.
  Appl. Cryst.}\ }\textbf {\bibinfo {volume} {50}},\ \bibinfo {pages} {1457}
  (\bibinfo {year} {2017})}\BibitemShut {NoStop}%
\bibitem [{\citenamefont {Bradlyn}\ \emph {et~al.}(2018)\citenamefont
  {Bradlyn}, \citenamefont {Elcoro}, \citenamefont {Vergniory}, \citenamefont
  {Wang}, \citenamefont {Cano}, \citenamefont {Felser}, \citenamefont {Aroyo},\
  and\ \citenamefont {Bernevig}}]{GraphTheoryPaper}%
  \BibitemOpen
  \bibfield  {author} {\bibinfo {author} {\bibfnamefont {B.}~\bibnamefont
  {Bradlyn}}, \bibinfo {author} {\bibfnamefont {L.}~\bibnamefont {Elcoro}},
  \bibinfo {author} {\bibfnamefont {M.~G.}\ \bibnamefont {Vergniory}}, \bibinfo
  {author} {\bibfnamefont {Z.}~\bibnamefont {Wang}}, \bibinfo {author}
  {\bibfnamefont {J.}~\bibnamefont {Cano}}, \bibinfo {author} {\bibfnamefont
  {C.}~\bibnamefont {Felser}}, \bibinfo {author} {\bibfnamefont {M.~I.}\
  \bibnamefont {Aroyo}}, \ and\ \bibinfo {author} {\bibfnamefont {B.~A.}\
  \bibnamefont {Bernevig}},\ }\href@noop {} {\bibfield  {journal} {\bibinfo
  {journal} {Phys. Rev. B}\ }\textbf {\bibinfo {volume} {97}},\ \bibinfo
  {pages} {035137} (\bibinfo {year} {2018})}\BibitemShut {NoStop}%
\bibitem [{\citenamefont {Song}\ \emph
  {et~al.}(2018{\natexlab{a}})\citenamefont {Song}, \citenamefont {Zhang},
  \citenamefont {Fang},\ and\ \citenamefont {Fang}}]{song2017}%
  \BibitemOpen
  \bibfield  {author} {\bibinfo {author} {\bibfnamefont {Z.}~\bibnamefont
  {Song}}, \bibinfo {author} {\bibfnamefont {T.}~\bibnamefont {Zhang}},
  \bibinfo {author} {\bibfnamefont {Z.}~\bibnamefont {Fang}}, \ and\ \bibinfo
  {author} {\bibfnamefont {C.}~\bibnamefont {Fang}},\ }\href@noop {} {\bibfield
   {journal} {\bibinfo  {journal} {Nature communications}\ }\textbf {\bibinfo
  {volume} {9}},\ \bibinfo {pages} {3530} (\bibinfo {year}
  {2018}{\natexlab{a}})}\BibitemShut {NoStop}%
\bibitem [{\citenamefont {Zhang}\ \emph {et~al.}(2018)\citenamefont {Zhang},
  \citenamefont {Jiang}, \citenamefont {Song}, \citenamefont {Huang},
  \citenamefont {He}, \citenamefont {Fang}, \citenamefont {Weng},\ and\
  \citenamefont {Fang}}]{bigmaterials-china}%
  \BibitemOpen
  \bibfield  {author} {\bibinfo {author} {\bibfnamefont {T.}~\bibnamefont
  {Zhang}}, \bibinfo {author} {\bibfnamefont {Y.}~\bibnamefont {Jiang}},
  \bibinfo {author} {\bibfnamefont {Z.}~\bibnamefont {Song}}, \bibinfo {author}
  {\bibfnamefont {H.}~\bibnamefont {Huang}}, \bibinfo {author} {\bibfnamefont
  {Y.}~\bibnamefont {He}}, \bibinfo {author} {\bibfnamefont {Z.}~\bibnamefont
  {Fang}}, \bibinfo {author} {\bibfnamefont {H.}~\bibnamefont {Weng}}, \ and\
  \bibinfo {author} {\bibfnamefont {C.}~\bibnamefont {Fang}},\ }\href@noop {}
  {\bibfield  {journal} {\bibinfo  {journal} {arXiv preprint arXiv:1807.08756}\
  } (\bibinfo {year} {2018})}\BibitemShut {NoStop}%
\bibitem [{\citenamefont {Cano}\ \emph {et~al.}(2018)\citenamefont {Cano},
  \citenamefont {Bradlyn}, \citenamefont {Wang}, \citenamefont {Elcoro},
  \citenamefont {Vergniory}, \citenamefont {Felser}, \citenamefont {Aroyo},\
  and\ \citenamefont {Bernevig}}]{EBRTheoryPaper}%
  \BibitemOpen
  \bibfield  {author} {\bibinfo {author} {\bibfnamefont {J.}~\bibnamefont
  {Cano}}, \bibinfo {author} {\bibfnamefont {B.}~\bibnamefont {Bradlyn}},
  \bibinfo {author} {\bibfnamefont {Z.}~\bibnamefont {Wang}}, \bibinfo {author}
  {\bibfnamefont {L.}~\bibnamefont {Elcoro}}, \bibinfo {author} {\bibfnamefont
  {M.~G.}\ \bibnamefont {Vergniory}}, \bibinfo {author} {\bibfnamefont
  {C.}~\bibnamefont {Felser}}, \bibinfo {author} {\bibfnamefont {M.~I.}\
  \bibnamefont {Aroyo}}, \ and\ \bibinfo {author} {\bibfnamefont {B.~A.}\
  \bibnamefont {Bernevig}},\ }\href@noop {} {\bibfield  {journal} {\bibinfo
  {journal} {Phys. Rev. B}\ }\textbf {\bibinfo {volume} {97}},\ \bibinfo
  {pages} {035139} (\bibinfo {year} {2018})}\BibitemShut {NoStop}%
\bibitem [{\citenamefont {Alexandradinata}\ and\ \citenamefont
  {H\"oller}(2018)}]{aris18-2}%
  \BibitemOpen
  \bibfield  {author} {\bibinfo {author} {\bibfnamefont {A.}~\bibnamefont
  {Alexandradinata}}\ and\ \bibinfo {author} {\bibfnamefont {J.}~\bibnamefont
  {H\"oller}},\ }\href {\doibase 10.1103/PhysRevB.98.184305} {\bibfield
  {journal} {\bibinfo  {journal} {Phys. Rev. B}\ }\textbf {\bibinfo {volume}
  {98}},\ \bibinfo {pages} {184305} (\bibinfo {year} {2018})}\BibitemShut
  {NoStop}%
\bibitem [{\citenamefont {Po}\ \emph {et~al.}(2018)\citenamefont {Po},
  \citenamefont {Watanabe},\ and\ \citenamefont {Vishwanath}}]{comment}%
  \BibitemOpen
  \bibfield  {author} {\bibinfo {author} {\bibfnamefont {H.~C.}\ \bibnamefont
  {Po}}, \bibinfo {author} {\bibfnamefont {H.}~\bibnamefont {Watanabe}}, \ and\
  \bibinfo {author} {\bibfnamefont {A.}~\bibnamefont {Vishwanath}},\
  }\href@noop {} {\bibfield  {journal} {\bibinfo  {journal} {Phys. Rev. Lett.}\
  }\textbf {\bibinfo {volume} {121}},\ \bibinfo {pages} {126402} (\bibinfo
  {year} {2018})}\BibitemShut {NoStop}%
\bibitem [{Note1()}]{Note1}%
  \BibitemOpen
  \bibinfo {note} {Because we will be considering in this work models both with
  and without time-reversal symmetry, we will follow Ref.~\protect
  \rev@citealpnum {Cracknell} and place a ``$1'$'' after a space group symbol
  when we consider time-reversal to be a symmetry of the system.}\BibitemShut
  {Stop}%
\bibitem [{\citenamefont {Bouhon}\ \emph {et~al.}(2018)\citenamefont {Bouhon},
  \citenamefont {Black-Schaffer},\ and\ \citenamefont {Slager}}]{Slager2018}%
  \BibitemOpen
  \bibfield  {author} {\bibinfo {author} {\bibfnamefont {A.}~\bibnamefont
  {Bouhon}}, \bibinfo {author} {\bibfnamefont {A.~M.}\ \bibnamefont
  {Black-Schaffer}}, \ and\ \bibinfo {author} {\bibfnamefont {R.-J.}\
  \bibnamefont {Slager}},\ }\href@noop {} {\  (\bibinfo {year} {2018})},\
  \Eprint {http://arxiv.org/abs/1804.09719} {arXiv:1804.09719} \BibitemShut
  {NoStop}%
\bibitem [{\citenamefont {{Cano}}\ \emph {et~al.}(2018)\citenamefont {{Cano}},
  \citenamefont {{Bradlyn}}, \citenamefont {{Wang}}, \citenamefont {{Elcoro}},
  \citenamefont {{Vergniory}}, \citenamefont {{Felser}}, \citenamefont
  {{Aroyo}},\ and\ \citenamefont {{Bernevig}}}]{Fragile2017}%
  \BibitemOpen
  \bibfield  {author} {\bibinfo {author} {\bibfnamefont {J.}~\bibnamefont
  {{Cano}}}, \bibinfo {author} {\bibfnamefont {B.}~\bibnamefont {{Bradlyn}}},
  \bibinfo {author} {\bibfnamefont {Z.}~\bibnamefont {{Wang}}}, \bibinfo
  {author} {\bibfnamefont {L.}~\bibnamefont {{Elcoro}}}, \bibinfo {author}
  {\bibfnamefont {M.~G.}\ \bibnamefont {{Vergniory}}}, \bibinfo {author}
  {\bibfnamefont {C.}~\bibnamefont {{Felser}}}, \bibinfo {author}
  {\bibfnamefont {M.~I.}\ \bibnamefont {{Aroyo}}}, \ and\ \bibinfo {author}
  {\bibfnamefont {B.~A.}\ \bibnamefont {{Bernevig}}},\ }\href@noop {}
  {\bibfield  {journal} {\bibinfo  {journal} {Phys. Rev. Lett.}\ }\textbf
  {\bibinfo {volume} {120}},\ \bibinfo {pages} {266401} (\bibinfo {year}
  {2018})}\BibitemShut {NoStop}%
\bibitem [{\citenamefont {Wieder}\ and\ \citenamefont
  {Bernevig}(2018)}]{wieder2018axion}%
  \BibitemOpen
  \bibfield  {author} {\bibinfo {author} {\bibfnamefont {B.~J.}\ \bibnamefont
  {Wieder}}\ and\ \bibinfo {author} {\bibfnamefont {B.~A.}\ \bibnamefont
  {Bernevig}},\ }\href@noop {} {\bibfield  {journal} {\bibinfo  {journal}
  {arXiv preprint arXiv:1810.02373}\ } (\bibinfo {year} {2018})}\BibitemShut
  {NoStop}%
\bibitem [{\citenamefont {Ahn}\ and\ \citenamefont
  {Yang}(2018)}]{ahn2018higher}%
  \BibitemOpen
  \bibfield  {author} {\bibinfo {author} {\bibfnamefont {J.}~\bibnamefont
  {Ahn}}\ and\ \bibinfo {author} {\bibfnamefont {B.-J.}\ \bibnamefont {Yang}},\
  }\href@noop {} {\bibfield  {journal} {\bibinfo  {journal} {arXiv preprint
  arXiv:1810.05363}\ } (\bibinfo {year} {2018})}\BibitemShut {NoStop}%
\bibitem [{\citenamefont {Benalcazar}\ \emph {et~al.}(2018)\citenamefont
  {Benalcazar}, \citenamefont {Li},\ and\ \citenamefont
  {Hughes}}]{benalcazar2018quantization}%
  \BibitemOpen
  \bibfield  {author} {\bibinfo {author} {\bibfnamefont {W.~A.}\ \bibnamefont
  {Benalcazar}}, \bibinfo {author} {\bibfnamefont {T.}~\bibnamefont {Li}}, \
  and\ \bibinfo {author} {\bibfnamefont {T.~L.}\ \bibnamefont {Hughes}},\
  }\href@noop {} {\bibfield  {journal} {\bibinfo  {journal} {arXiv preprint
  arXiv:1809.02142}\ } (\bibinfo {year} {2018})}\BibitemShut {NoStop}%
\bibitem [{\citenamefont {Legner}\ and\ \citenamefont
  {Neupert}(2013)}]{Legner2013}%
  \BibitemOpen
  \bibfield  {author} {\bibinfo {author} {\bibfnamefont {M.}~\bibnamefont
  {Legner}}\ and\ \bibinfo {author} {\bibfnamefont {T.}~\bibnamefont
  {Neupert}},\ }\href@noop {} {\bibfield  {journal} {\bibinfo  {journal} {Phys.
  Rev. B}\ }\textbf {\bibinfo {volume} {88}},\ \bibinfo {pages} {115114}
  (\bibinfo {year} {2013})}\BibitemShut {NoStop}%
\bibitem [{\citenamefont {L{\"o}wdin}(1950)}]{Lowdin1950}%
  \BibitemOpen
  \bibfield  {author} {\bibinfo {author} {\bibfnamefont {P.-O.}\ \bibnamefont
  {L{\"o}wdin}},\ }\href@noop {} {\bibfield  {journal} {\bibinfo  {journal}
  {The Journal of Chemical Physics}\ }\textbf {\bibinfo {volume} {18}},\
  \bibinfo {pages} {365} (\bibinfo {year} {1950})}\BibitemShut {NoStop}%
\bibitem [{Note2()}]{Note2}%
  \BibitemOpen
  \bibinfo {note} {See for instance Eq.~(S43) of Ref.~\protect \rev@citealpnum
  {NaturePaper}, and also Footnote $34$ of Ref.~\protect \rev@citealpnum
  {Soluyanov2011}}\BibitemShut {NoStop}%
\bibitem [{\citenamefont {Alexandradinata}\ \emph {et~al.}(2014)\citenamefont
  {Alexandradinata}, \citenamefont {Fang}, \citenamefont {Gilbert},\ and\
  \citenamefont {Bernevig}}]{Alexandradinata14}%
  \BibitemOpen
  \bibfield  {author} {\bibinfo {author} {\bibfnamefont {A.}~\bibnamefont
  {Alexandradinata}}, \bibinfo {author} {\bibfnamefont {C.}~\bibnamefont
  {Fang}}, \bibinfo {author} {\bibfnamefont {M.~J.}\ \bibnamefont {Gilbert}}, \
  and\ \bibinfo {author} {\bibfnamefont {B.~A.}\ \bibnamefont {Bernevig}},\
  }\href {\doibase 10.1103/PhysRevLett.113.116403} {\bibfield  {journal}
  {\bibinfo  {journal} {Phys. Rev. Lett.}\ }\textbf {\bibinfo {volume} {113}},\
  \bibinfo {pages} {116403} (\bibinfo {year} {2014})}\BibitemShut {NoStop}%
\bibitem [{\citenamefont {Alexandradinata}\ \emph {et~al.}(2016)\citenamefont
  {Alexandradinata}, \citenamefont {Wang},\ and\ \citenamefont
  {Bernevig}}]{ArisCohomology}%
  \BibitemOpen
  \bibfield  {author} {\bibinfo {author} {\bibfnamefont {A.}~\bibnamefont
  {Alexandradinata}}, \bibinfo {author} {\bibfnamefont {Z.}~\bibnamefont
  {Wang}}, \ and\ \bibinfo {author} {\bibfnamefont {B.~A.}\ \bibnamefont
  {Bernevig}},\ }\href {\doibase 10.1103/PhysRevX.6.021008} {\bibfield
  {journal} {\bibinfo  {journal} {Phys. Rev. X}\ }\textbf {\bibinfo {volume}
  {6}},\ \bibinfo {pages} {021008} (\bibinfo {year} {2016})}\BibitemShut
  {NoStop}%
\bibitem [{\citenamefont {Vergniory}\ \emph {et~al.}(2017)\citenamefont
  {Vergniory}, \citenamefont {Elcoro}, \citenamefont {Wang}, \citenamefont
  {Cano}, \citenamefont {Felser}, \citenamefont {Aroyo}, \citenamefont
  {Bernevig},\ and\ \citenamefont {Bradlyn}}]{GraphDataPaper}%
  \BibitemOpen
  \bibfield  {author} {\bibinfo {author} {\bibfnamefont {M.~G.}\ \bibnamefont
  {Vergniory}}, \bibinfo {author} {\bibfnamefont {L.}~\bibnamefont {Elcoro}},
  \bibinfo {author} {\bibfnamefont {Z.}~\bibnamefont {Wang}}, \bibinfo {author}
  {\bibfnamefont {J.}~\bibnamefont {Cano}}, \bibinfo {author} {\bibfnamefont
  {C.}~\bibnamefont {Felser}}, \bibinfo {author} {\bibfnamefont {M.~I.}\
  \bibnamefont {Aroyo}}, \bibinfo {author} {\bibfnamefont {B.~A.}\ \bibnamefont
  {Bernevig}}, \ and\ \bibinfo {author} {\bibfnamefont {B.}~\bibnamefont
  {Bradlyn}},\ }\href@noop {} {\bibfield  {journal} {\bibinfo  {journal} {Phys.
  Rev. E}\ }\textbf {\bibinfo {volume} {96}},\ \bibinfo {pages} {023310}
  (\bibinfo {year} {2017})}\BibitemShut {NoStop}%
\bibitem [{\citenamefont {Kane}\ and\ \citenamefont {Mele}(2005)}]{Kane04}%
  \BibitemOpen
  \bibfield  {author} {\bibinfo {author} {\bibfnamefont {C.~L.}\ \bibnamefont
  {Kane}}\ and\ \bibinfo {author} {\bibfnamefont {E.~J.}\ \bibnamefont
  {Mele}},\ }\href@noop {} {\bibfield  {journal} {\bibinfo  {journal} {Phys.
  Rev. Lett.}\ }\textbf {\bibinfo {volume} {95}},\ \bibinfo {pages} {226801}
  (\bibinfo {year} {2005})}\BibitemShut {NoStop}%
\bibitem [{\citenamefont {Vergniory}\ \emph {et~al.}(2018)\citenamefont
  {Vergniory}, \citenamefont {Elcoro}, \citenamefont {Felser}, \citenamefont
  {Bernevig},\ and\ \citenamefont {Wang}}]{bigmaterials}%
  \BibitemOpen
  \bibfield  {author} {\bibinfo {author} {\bibfnamefont {M.}~\bibnamefont
  {Vergniory}}, \bibinfo {author} {\bibfnamefont {L.}~\bibnamefont {Elcoro}},
  \bibinfo {author} {\bibfnamefont {C.}~\bibnamefont {Felser}}, \bibinfo
  {author} {\bibfnamefont {B.}~\bibnamefont {Bernevig}}, \ and\ \bibinfo
  {author} {\bibfnamefont {Z.}~\bibnamefont {Wang}},\ }\href@noop {} {\bibfield
   {journal} {\bibinfo  {journal} {arXiv preprint arXiv:1807.10271}\ }
  (\bibinfo {year} {2018})}\BibitemShut {NoStop}%
\bibitem [{\citenamefont {Schindler}\ \emph {et~al.}(2018)\citenamefont
  {Schindler}, \citenamefont {Wang}, \citenamefont {Vergniory}, \citenamefont
  {Cook}, \citenamefont {Murani}, \citenamefont {Sengupta}, \citenamefont
  {Kasumov}, \citenamefont {Deblock}, \citenamefont {Jeon}, \citenamefont
  {Drozdov} \emph {et~al.}}]{schindler2018higher}%
  \BibitemOpen
  \bibfield  {author} {\bibinfo {author} {\bibfnamefont {F.}~\bibnamefont
  {Schindler}}, \bibinfo {author} {\bibfnamefont {Z.}~\bibnamefont {Wang}},
  \bibinfo {author} {\bibfnamefont {M.~G.}\ \bibnamefont {Vergniory}}, \bibinfo
  {author} {\bibfnamefont {A.~M.}\ \bibnamefont {Cook}}, \bibinfo {author}
  {\bibfnamefont {A.}~\bibnamefont {Murani}}, \bibinfo {author} {\bibfnamefont
  {S.}~\bibnamefont {Sengupta}}, \bibinfo {author} {\bibfnamefont {A.~Y.}\
  \bibnamefont {Kasumov}}, \bibinfo {author} {\bibfnamefont {R.}~\bibnamefont
  {Deblock}}, \bibinfo {author} {\bibfnamefont {S.}~\bibnamefont {Jeon}},
  \bibinfo {author} {\bibfnamefont {I.}~\bibnamefont {Drozdov}},  \emph
  {et~al.},\ }\href@noop {} {\bibfield  {journal} {\bibinfo  {journal} {Nature
  Physics}\ }\textbf {\bibinfo {volume} {14}},\ \bibinfo {pages} {918}
  (\bibinfo {year} {2018})}\BibitemShut {NoStop}%
\bibitem [{\citenamefont {Liu}\ \emph {et~al.}(2018)\citenamefont {Liu},
  \citenamefont {Emmanouilidou}, \citenamefont {Xing}, \citenamefont {Graf},\
  and\ \citenamefont {Ni}}]{liu2018quantum}%
  \BibitemOpen
  \bibfield  {author} {\bibinfo {author} {\bibfnamefont {J.}~\bibnamefont
  {Liu}}, \bibinfo {author} {\bibfnamefont {E.}~\bibnamefont {Emmanouilidou}},
  \bibinfo {author} {\bibfnamefont {J.}~\bibnamefont {Xing}}, \bibinfo {author}
  {\bibfnamefont {D.}~\bibnamefont {Graf}}, \ and\ \bibinfo {author}
  {\bibfnamefont {N.}~\bibnamefont {Ni}},\ }\href@noop {} {\bibfield  {journal}
  {\bibinfo  {journal} {arXiv preprint arXiv:1807.02546}\ } (\bibinfo {year}
  {2018})}\BibitemShut {NoStop}%
\bibitem [{\citenamefont {Song}\ \emph
  {et~al.}(2018{\natexlab{b}})\citenamefont {Song}, \citenamefont {Wang},
  \citenamefont {Shi}, \citenamefont {Li}, \citenamefont {Fang},\ and\
  \citenamefont {Bernevig}}]{zhidaprep}%
  \BibitemOpen
  \bibfield  {author} {\bibinfo {author} {\bibfnamefont {Z.}~\bibnamefont
  {Song}}, \bibinfo {author} {\bibfnamefont {Z.}~\bibnamefont {Wang}}, \bibinfo
  {author} {\bibfnamefont {W.}~\bibnamefont {Shi}}, \bibinfo {author}
  {\bibfnamefont {G.}~\bibnamefont {Li}}, \bibinfo {author} {\bibfnamefont
  {C.}~\bibnamefont {Fang}}, \ and\ \bibinfo {author} {\bibfnamefont {B.~A.}\
  \bibnamefont {Bernevig}},\ }\href@noop {} {\bibfield  {journal} {\bibinfo
  {journal} {arXiv preprint arXiv:1807.10676}\ } (\bibinfo {year}
  {2018}{\natexlab{b}})}\BibitemShut {NoStop}%
\bibitem [{\citenamefont {Fang}\ \emph {et~al.}(2012)\citenamefont {Fang},
  \citenamefont {Gilbert},\ and\ \citenamefont {Bernevig}}]{Fang2012}%
  \BibitemOpen
  \bibfield  {author} {\bibinfo {author} {\bibfnamefont {C.}~\bibnamefont
  {Fang}}, \bibinfo {author} {\bibfnamefont {M.~J.}\ \bibnamefont {Gilbert}}, \
  and\ \bibinfo {author} {\bibfnamefont {B.~A.}\ \bibnamefont {Bernevig}},\
  }\href {\doibase 10.1103/PhysRevB.86.115112} {\bibfield  {journal} {\bibinfo
  {journal} {Phys. Rev. B}\ }\textbf {\bibinfo {volume} {86}},\ \bibinfo
  {pages} {115112} (\bibinfo {year} {2012})}\BibitemShut {NoStop}%
\bibitem [{\citenamefont {Bradley}\ and\ \citenamefont
  {Cracknell}(1972)}]{Cracknell}%
  \BibitemOpen
  \bibfield  {author} {\bibinfo {author} {\bibfnamefont {C.~J.}\ \bibnamefont
  {Bradley}}\ and\ \bibinfo {author} {\bibfnamefont {A.~P.}\ \bibnamefont
  {Cracknell}},\ }\href@noop {} {\emph {\bibinfo {title} {The Mathematical
  Theory of Symmetry in Solids}}}\ (\bibinfo  {publisher} {Clarendon Press},\
  \bibinfo {address} {Oxford},\ \bibinfo {year} {1972})\BibitemShut {NoStop}%
\bibitem [{\citenamefont {Soluyanov}\ and\ \citenamefont
  {Vanderbilt}(2011)}]{Soluyanov2011}%
  \BibitemOpen
  \bibfield  {author} {\bibinfo {author} {\bibfnamefont {A.~A.}\ \bibnamefont
  {Soluyanov}}\ and\ \bibinfo {author} {\bibfnamefont {D.}~\bibnamefont
  {Vanderbilt}},\ }\href {\doibase 10.1103/PhysRevB.83.035108} {\bibfield
  {journal} {\bibinfo  {journal} {Phys. Rev. B}\ }\textbf {\bibinfo {volume}
  {83}},\ \bibinfo {pages} {035108} (\bibinfo {year} {2011})}\BibitemShut
  {NoStop}%
\end{thebibliography}%
\end{document}